\documentclass[preprint, number, 12pt]{elsarticle}

\usepackage{graphicx}
\usepackage{url}
\usepackage{amsmath}
\usepackage{amssymb}

\journal{Physica D}

\begin{document}

\begin{frontmatter}



\title{Sustained turbulence in the three-dimensional Gross-Pitaevskii model}


\author[label1,label2]{Davide Proment}
\address[label1]{Dipartimento di Fisica Generale, Universit\`{a} di Torino, Via Pietro Giuria 1, 10125 Torino, Italy}
\address[label2]{INFN, Sezione di Torino, Via Pietro Giuria 1, 10125 Torino, Italy}

\author[label3]{Sergey Nazarenko}
\address[label3]{Mathematics Institute, The University of Warwick, Coventry, CV4-7AL, UK}

\author[label1]{Miguel Onorato}

\begin{abstract}
We study the 3D forced-dissipated Gross-Pitaevskii equation.
We force  at relatively low wave numbers, expecting to observe a direct
energy cascade and a consequent power-law spectrum of the form $k^{-\alpha}$.
Our numerical results show that the exponent $\alpha$ strongly depends on
how the inverse particle cascade is attenuated at  $k$'s lower than
the forcing wave number. If the inverse cascade is arrested
by a friction at low $k$'s, we observe an exponent which is in good agreement with the
weak wave turbulence prediction $k^{-1}$. For a hypo-viscosity, a  $k^{-2}$
spectrum is observed which we explain using a critical balance argument.
In  simulations without any low-$k$ dissipation,
a condensate at $k=0$ is growing
and the system  goes through a strongly-turbulent transition from a four-wave to a three-wave
weak turbulence acoustic regime with $k^{-3/2}$ Zakharov-Sagdeev spectrum.
In this regime, we also observe a spectrum for the incompressible kinetic energy
which formally resembles the  Kolmogorov  $k^{-5/3}$, but whose correct explanation
should be in terms of the Kelvin wave turbulence. The probability density functions
for the velocities and the densities  are also discussed.
\end{abstract}


\end{frontmatter}

\section{INTRODUCTION} \label{intro}
After the seminal papers by A. Kolmogorov in 1941, it is well established that, apart from small corrections due to intermittency
\cite{frisch1995t},
the energy spectrum, $E(k)$, of the velocity fluctuations for high Reynolds number hydrodynamic turbulence shows a power  law
of the form $E(k)=C\;  P^{2/3} \;  k^{-5/3}$, where $C$ is the dimensionless Kolmogorov  constant and $P$ is
the flux of energy in the wave number space. This is a very strong result that has been  confirmed experimentally and
  numerically by the direct numerical integration of the Navier-Stokes equation. It can be  obtained
via dimensional considerations or as a solution of phenomenological turbulence
closures \cite{frisch1995t}. However, so far, this result has not been obtained
analytically from the Navier-Stokes equation.
Many decades after the work by Kolmogorov,
it has been discovered by Zakharov in 1965
that systems of weakly nonlinear, dispersive, random waves
behave qualitatively in a similar way as hydrodynamical turbulence \cite{zakh65}.
Namely,
the nonlinear interaction of waves can produce other waves with different
wavelengths and so on, generating a {\it cascade} process
leading to power-law wave spectra similar to the Kolmogorov spectrum.
Because of such an analogy, they  are called Kolmogorov-Zakharov (KZ) spectra, and the entire field
is known as Weak Wave Turbulence (WWT).
Description of WWT turns out to be more accessible
than of the hydrodynamic turbulence because nonlinearity
 in the dynamical equations, although still crucial, is small.
 In the WWT framework, a systematic approach
based on averaging the dynamical equations leads to
a Boltzmann-like  equation known as the
{\it wave kinetic equation}
 which describes the evolution of the spectrum of
  the turbulent wave field \cite{zakharov41kst}.
One remarkable property of WWT is that, in
contrast to hydrodynamic turbulence, the KZ spectra have been found as
exact stationary solutions of the wave kinetic equation \cite{zakh65,zakharov1967ess}.
Unlike the thermodynamic solutions,
for which  the integrand in the collision integral is identically zero,
the KZ solutions
correspond to non-trivial
states for which
a source, a sink and a window of transparency (inertial range) are required.
 Since this discovery, weak wave turbulence has found applications for a
 vast variety of  physical systems ranging from quantum to astrophysical scales,
 see books \cite{zakharov41kst,nazarenko-book} and references therein.

In this paper, we consider nonlinear dispersive  waves described by Gross-Pitaevskii equation (GPE),
%
\begin{equation}
i\frac{\partial \psi}{\partial t}+\nabla^2\psi +\sigma |\psi|^2\psi=0.
\label{nls0}
\end{equation}
This partial differential equation
has attracted the attention of many researchers:
it describes
propagation of optical pulses in  nonlinear media \cite{sulem1899nse} and
weakly interacting boson gases at very low temperatures called Bose-Einstein condensates
(BEC) \cite{dalfovo1999theory}.
In the present paper we will be concerned with the latter case and we will focus on the three-dimensional (3D) systems.
  Complex wave function
$ \psi $  is called {\it order parameter} and $ \sigma = \pm1$, depending on the physics of the problem:
  the defocusing case, $ \sigma=-1$, represents repulsive bosons, while the focusing, $ \sigma=1 $, considers an attractive
  interaction.
GPE has recently attracted the attention of many fluid dynamicists because BEC is a good example of superfluid,
i.e. a fluid with zero viscosity.
Indeed, using the Madelung transformation, the GPE can be mapped onto an Euler equation
which differs from the classical one only by an extra
term named  quantum pressure. Therefore, numerical computations of the 3D defocusing GPE
 have become a tool for investigating superfluids and quantum fluids dynamics.
Phenomena such as the vortex reconnections \cite{koplik1993vrs}, formation of a condensate \cite{berloff2002ssn},
and formation of power-law spectra  \cite{nore1997dkt, kobayashi2005kss, kobayashi2005kol, proment:051603} have been observed.

The purpose of this paper is to revisit and investigate the turbulence characteritics in the defocusing 3D GPE
with particular attention to the forced and dissipated case. An interesting issue to be addressed is
verification of the  WWT theory which offers a solid theoretical tool for predicting statistical
quantities in systems of dispersive, weakly interacting waves. As it will be shown
in the next section, when the nonlinearity becomes large, the predictions of the WWT theory fail
 and the concept of  {\it critical balance} (CB) has to be introduced in order to
explain the observed spectra.

This paper is organized as follows. In section \ref{theory} we present the theoretical background
 on the GPE model revisiting some mathematical aspects, including  quantum vortices and the
 general properties of quantum turbulence. In section \ref{fd-GPE} we introduce the
 forced-dissipated GPE and present predictions for steady turbulent states: in
 particular we discuss the WWT for the  four-wave and the three-wave regimes
  and the CB conjecture. Section \ref{numerics} is dedicated to presenting numerical
  results in three different regimes: free condensation at large scales ({\bf RUN 1}),
  dissipation  by friction at low wave numbers ({\bf RUN 2}),
  and dissipation  by hypo-viscosity at low wave numbers ({\bf RUN 3}).
  Finally, section \ref{conclusion} contains the conclusions.

\section{Theoretical background} \label{theory}
In this paper, we will consider the defocusing GPE model,
\begin{equation}
\imath \frac{\partial \psi(\mathbf{x}, t)}{\partial t} + \nabla^2 \psi(\mathbf{x}, t) - |\psi(\mathbf{x}, t)|^2\psi(\mathbf{x}, t)=0,
\label{gpe}
\end{equation}
where the nonlinearity comes from the self-interactions proportional  to the gas density $ \rho(\mathbf{x}, t)=|\psi(\mathbf{x}, t)|^2 $;
this term is a consequence of considering local interactions between bosons.
The system is conservative and its Hamiltonian is
\begin{equation}
H=\int \left( |\nabla \psi(\mathbf{x}, t)|^2 + \frac{1}{2}|\psi(\mathbf{x}, t)|^4 \right) d\mathbf{x} = H_{lin}(t) + H_{nl}(t).
\end{equation}
In the latter relation the total energy $ H $ has been split into a part responsible for the linear dynamics, $ H_{lin}(t) = \int |\nabla\psi(\mathbf{x}, t)|^2 d\mathbf{k} $, and the one describing the  nonlinear interactions, $ H_{nl}(t) =1/2 \int |\psi(\mathbf{x}, t)|^4 d\mathbf{k} $.
In the following, we will use the  energy densities defined as  $ \mathcal{E}_{lin}=H_{lin}/V $ and $ \mathcal{E}_{nl}=H_{nl}/V $ where $ V $ is the total volume.
The system conserves a second quantity, the mass $ M $, defined as
\begin{equation}
M=\int {|\psi(\mathbf{x}, t)|^2} d\mathbf{x} = \int \rho(\mathbf{x}, t) d\mathbf{x}.
\end{equation}

An important quantity that characterizes the system is the healing length $ \xi $: it physically  estimates the distance over which the field $ \psi(\mathbf{x}, t) $ recovers its bulk value when subject to a localized perturbation. This definition refers to the case of a single perturbation in a uniform field but can be extended to even  highly perturbed statistical  systems as
\begin{equation}
\xi = \frac{1}{\sqrt{\langle \rho \rangle}},
\end{equation}
where $ \langle \cdot \rangle $ denotes the spatial averaging, i.e. $ \langle \rho \rangle = M/V $.  The healing length $ \xi $ measures on average the scale at which the nonlinear term becomes comparable with the linear one. Dual to  the scale $\xi$ is  wave number
$ k_{\xi}=2\pi \sqrt{\langle \rho \rangle} $.

The GPE has been widely studied in the fluid dynamics framework. Indeed, the  Madelung transformation, $ \psi(\mathbf{x}, t)=\sqrt{\rho(\mathbf{x}, t)}e^{\imath \theta(\mathbf{x}, t)}$, maps the GPE for the complex field  into the system of two equations,
\begin{equation}
\begin{split}
& \frac{\partial \rho}{\partial t} + \nabla \cdot (\rho \mathbf{v}) = 0 \\
&\rho \left( \frac{\partial v_j}{\partial t} + v_k \frac{\partial v_j}{\partial x_k} \right) = -\frac{\partial p}{\partial x_j} + \frac{\partial \Sigma_{jk}}{\partial x_k},
\label{eq:fd-NLSE}
\end{split}
\end{equation}
for the real density field $ \rho(\mathbf{x}, t) $ and a real velocity field $ \mathbf{v}(\mathbf{x}, t) = 2\nabla\theta(\mathbf{x}, t) $.
The first equation represents a continuity equation for a compressible fluid and the second equation is a momentum conservation law.
 The  terms in the r.h.s of the latter equation can be thought as a pressure $ p=\rho^2 $ and a ``quantum stress" tensor, $\Sigma_{jk}= \rho \frac{\partial^2  (\ln \rho)}{\partial x_j \partial x_k} $. The quantum stress term
 becomes important at scales of the order of $ \xi $.
 The system (\ref{eq:fd-NLSE}) describes an inviscid and irrotational fluid flow.

\subsection{Quantum vortices}
\begin{figure}
\begin{minipage}[l]{0.36\linewidth}
\includegraphics[scale=0.5]{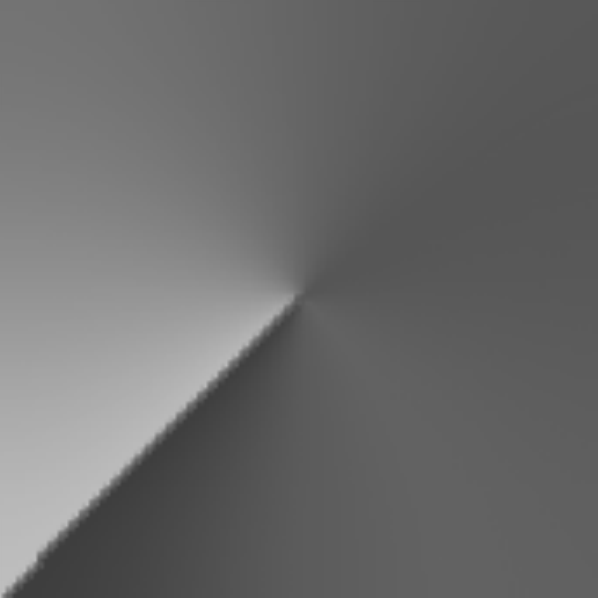}
\caption{$ \theta(\mathbf{x}, t) $ near a 2D vortex. \label{vortex-phase}}
\includegraphics[scale=0.48]{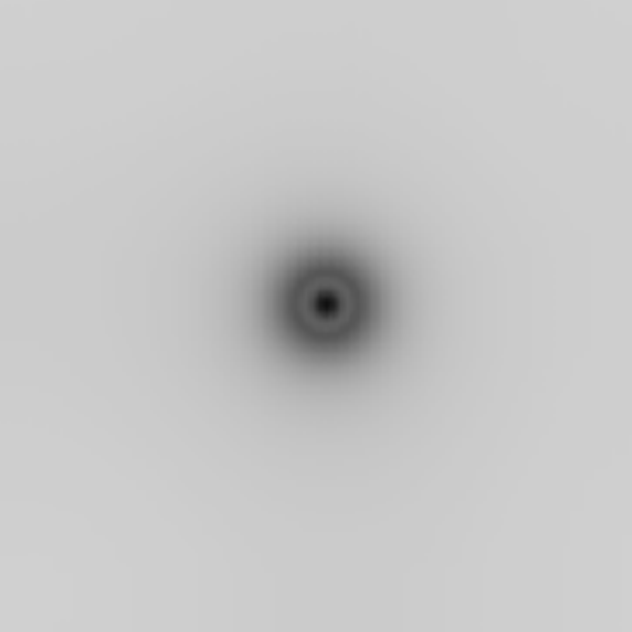}
\caption{$ \rho(\mathbf{x}, t) $ near a 2D vortex. \label{vortex-density}}
\end{minipage}
\begin{minipage}[r]{0.58\linewidth}
\includegraphics[scale=0.28]{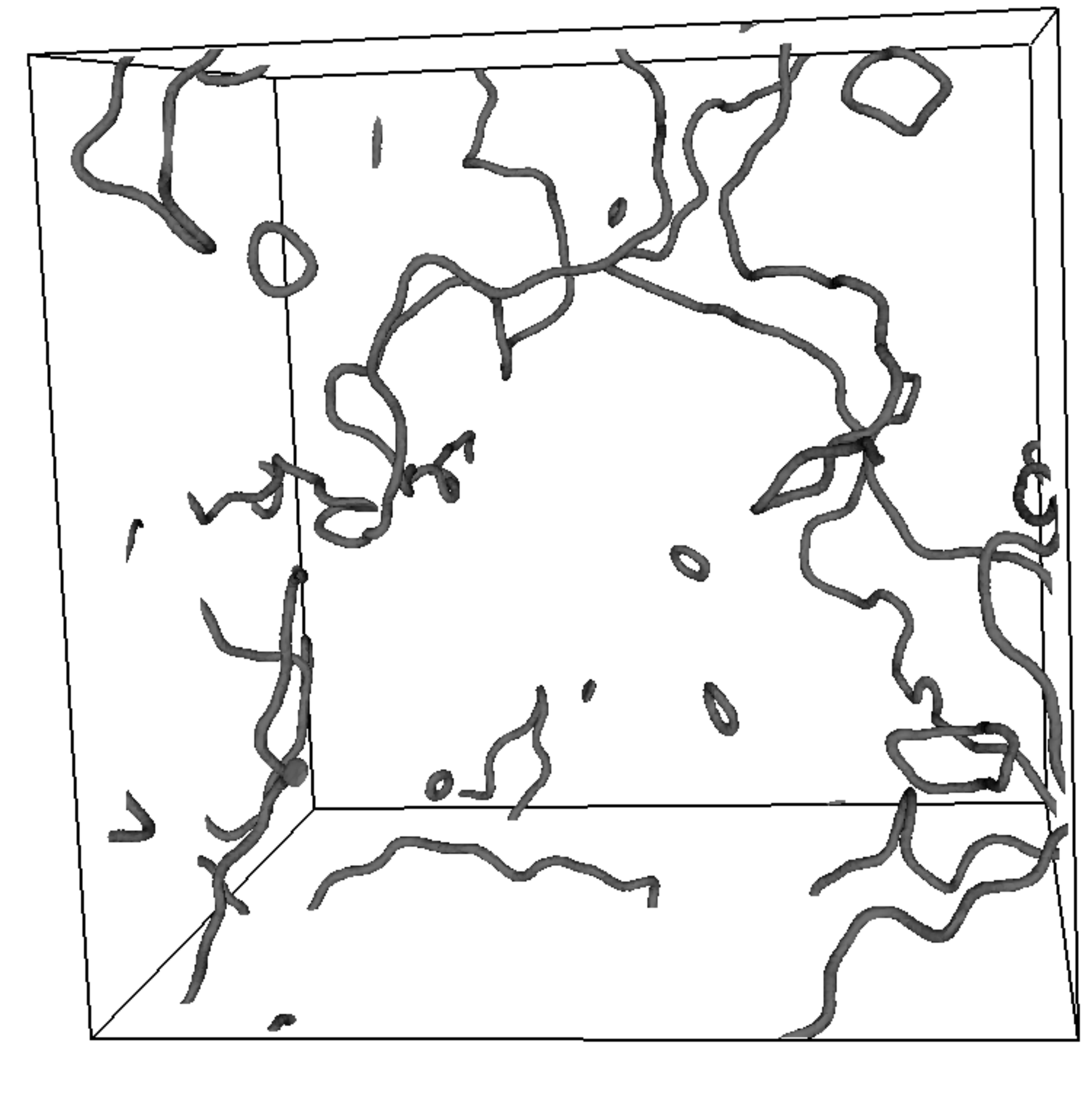}
\caption{Plot of the low density region $ \rho(\mathbf{x}, t) $ below a certain threshold in a freely decay simulation. \label{vortices}}
\end{minipage}
\end{figure}

Even if the fluid is irrotational, particular types of vortex solutions exist. This is true  if the region occupied by
 the irrotational flow is not simply connected, e.g., if  there are phase defects in the field
 at locations of the zero density.
Moreover, differently from classical fluids, such vortices carry quantized circulation and therefore are called {\it quantum vortices}.
To better understand  their structure, we first consider a two-dimensional (2D)
system. A necessary condition to assure the continuity of the complex field $ \psi $ is that the phase changes by $ \Delta \theta = 2\pi n $, where $ n \in \mathcal{N} $, around a vortex. As the velocity field $ \mathbf{v} $ is proportional to the gradient of the phase, it is easy to see that the circulation around a vortex is quantized,
\begin{equation}
\mathcal{C} = \oint \mathbf{v} \cdot d\mathbf{l} = 2 \oint \nabla \theta \cdot d\mathbf{l} = 2 \Delta \theta.
\end{equation}
In Fig. \ref{vortex-phase} and Fig. \ref{vortex-density} we show respectively the phase field $ \theta(\mathbf{x}, t) $ and the density field $ \rho(\mathbf{x}, t) $ in the neighborhood of a 2D vortex embedded in a uniform density field.
The vortex core size is of the order of $ \xi $ because, by definition, the healing length  measures the size of a generic order-one
fluctuation on the uniform condensate solution. By measuring circulation it is possible to distinguish between
clockwise and anti-clockwise vortices.
Note another important difference from the classical fluids arising from the presence of the
quantum stress: vortices with the opposite sign can approach each other and  annihilate.

In 3D, the vortices are more complicated objects consisting of continuous lines or loops with different topologies. These structures can oscillate producing Kelvin waves \cite{kozik:060502}, be transported by the fluid or induce a fluid motion, and reconnect \cite{koplik1993vrs, alamri:215302}. The vortex energy is proportional, in the leading order, to its length and it can be transferred to the fluid via sound waves \cite{leadbeater2001sou}. Examples of all these vortex motions is shown in Fig. \ref{vortices}: here we plot a snapshot of low density regions,
associated with the vortex cores, obtained in a numerical computation of the GPE in a periodic cubic box with
twelve vortices as initial condition \cite{url:proment-qt}.

\subsection{Quantum Turbulence}
Vortex-vortex dynamics (collective dynamics of vortex bundles, reconnections of vortex lines), vortex-sound (radiation and scattering of sound by vortices) and sound-sound  interactions (acoustic wave turbulence) and the dynamics of the vortex itself (Kelvin waves, vortex rings) represent the ingredients of quantum turbulence (QT).
When forced at the scales larger than the mean inter-vortex separation $\ell \gg \xi$, QT cascade starts as the classical
Richardson cascade characteristic of Navier-Stokes turbulence until it reaches the scale  $\ell$. At scales $\le \ell$ the
discreteness of the quantized vortex field becomes essential, and the further cascade to lower scales is very different than the
one in the classical fluids with continuous vorticity  \cite{vinen-nimela, PhysRevB.61.1410, PhysRevB.64.134520}. After some reconnections and other crossover processes
 near the transitional  scale  $\ell$ \cite{lvov_naz_rud1,lvov_naz_rud2},
the energy cascade is be carried to the scales  $\le \ell$ by Kelvin wave cascade until,
 at a very low scale, it is radiated into sound
 \cite{vinen-nimela, PhysRevB.61.1410, PhysRevB.64.134520, PhysRevLett.92.035301, boffetta, KW_rec, KW-dam, LN-spectrum}.
 
In the recent years, GPE has become a popular model for studying QT by using numerical simulations \cite{nore1997dkt, berloff2002ssn, kobayashi2005kss, kobayashi2005kol, yepez:084501, PhysRevLett.104.075301}. The pioneering paper of Nore et al. \cite{nore1997dkt} showed similarities between the 3D GPE and the classical Navier-Stokes turbulence, including observations of the famous Kolmogorov $5/3$-spectrum.
This spectrum was observed for scales $>\ell $ (with $\ell > \xi$), which agrees with the view that at this scales the discreteness of the vortex lines is inessential, and the turbulence picture is basically classical Kolmogorov.
However, in several follow-up works  \cite{kobayashi2005kss, kobayashi2005kol, yepez:084501} the $5/3$-spectrum was reported to extend
to the smaller scales, between $\ell $ and $\xi$, where the classical Kolmogorov picture is not expected to be valid theoretically.
Later in our paper we will comment on this apparent paradox.

 Numerical quantities usually measured in QT can be defined in the fluid dynamic framework. Namely, the
 total energy is divided into kinetic, quantum and internal energy of the system as follows \cite{nore1997dkt},
\begin{equation}
H = E_{kin}(t) + E_{qu}(t) + E_{int}(t),
\end{equation}
where $ E_{kin}(t) = \frac{1}{2} \int \left( \sqrt{\rho} \mathbf{v} \right)^2 d\mathbf{x} $ is the kinetic energy , $ E_{qu}(t) = \frac{1}{2} \int \left(2\nabla\sqrt{\rho} \right)^2 d\mathbf{x} $ is the quantum energy
 and  $ E_{int}(t) = \frac{1}{2} \int  \rho^2 d\mathbf{x} = H_{nl}(t) $ the internal energy.
Moreover, the quantity in the integrand of the kinetic energy is usually divided, using the Helmholtz's theorem, into a solenoidal (incompressible) and an irrotational (compressible) vector fields,
\begin{equation}
\sqrt{\rho}\mathbf{v} =  (\sqrt{\rho}\mathbf{v})^i + (\sqrt{\rho}\mathbf{v})^c.
\end{equation}
The incompressible part, satisfying  $ \nabla \cdot (\sqrt{\rho}\mathbf{v})^i=0 $, is associated to the vortex dynamics; the compressible part, which satisfies $ \nabla \times (\sqrt{\rho}\mathbf{v})^c=0 $, is related to the  sound waves. Note that the mentioned
 above results of  papers \cite{nore1997dkt, kobayashi2005kss, kobayashi2005kol, yepez:084501} for the $ k^{-5/3} $ power-law behavior
refer to the spectrum of the incompressible kinetic energy $ E_{kin}^i $.


\section{Forced-dissipated GPE}\label{fd-GPE}
The aim of this research is to study the steady turbulent states in the GPE model, with particular interest in observing cascades from large scales to small ones. Therefore, in the spirit of classical turbulence, we build a sustained system by including a forcing and a dissipation terms in  the GPE:
\begin{equation}
\imath \frac{\partial \psi(\mathbf{x}, t)}{\partial t} + \nabla^2 \psi(\mathbf{x}, t) - |\psi(\mathbf{x}, t)|^2\psi(\mathbf{x}, t)= \mathcal{F} + \mathcal{D},
\label{gpe-fd}
\end{equation}
where the forcing $ \mathcal{F} $ injects mass and energy, while the dissipation $ \mathcal{D} $ removes them. As a consequence, the mass and the energy are no longer constants of motion. However, if a steady turbulent state is reached, the quantities $ \dot{H}(t) $ and $ \dot{M}(t) $ are zero.
In such a steady state, the mass and the energy get transferred  from the forcing to the dissipation scales, being approximately conserved
while cascading though the inertial ranges (see  \ref{ap:fluxes}).

\subsection{Weak wave turbulence predictions}
The WWT furnishes some prediction for wave spectra in steady states. In order to
introduce WWT, let us put the system into a 3D periodic box and write the equation (neglecting $ \mathcal{F} $ and $ \mathcal{D} $) in Fourier space:
\begin{equation}
\imath \frac{\partial \tilde{\psi}(\mathbf{k}_1, t)}{\partial t} - \omega(\mathbf{k}_1) \tilde{\psi}(\mathbf{k}_1, t) =
\sum_{\mathbf{k}_{2}, \mathbf{k}_{3}, \mathbf{k}_{4} \in \mathbb{Z}^3}
 \tilde{\psi}^\ast(\mathbf{k}_2, t) \tilde{\psi}(\mathbf{k}_3, t) \tilde{\psi}(\mathbf{k}_4, t) \delta(\mathbf{k}_1 + \mathbf{k}_2 -\mathbf{k}_3-\mathbf{k}_4) .
\label{eq:bbgky}
\end{equation}
Here $\tilde{\psi}_i \equiv  \tilde{\psi}(\mathbf{k}_i, t) $ is the Fourier transform
of ${\psi}(\mathbf{x}, t) $ and $\omega(\mathbf{k})={k}^2 $ is the dispersion relation of the system
(hereafter $k=|\mathbf{k}|$).
The WWT theory, taking the limits of infinite-box limit, small nonlinearity,
and assuming space homogeneity and
random  phases and amplitudes (RPA) of the initial  wave amplitudes,
provides a statistical closure.
The closure predicts the behavior of statistical quantities such as correlators $ \langle \tilde{\psi}_1 \tilde{\psi}_2 ... \tilde{\psi}_n \tilde{\psi}_{n+1}^\ast \tilde{\psi}_{n+2}^\ast ... \tilde{\psi}_{n+m}^\ast \rangle  $, where the average is performed over the random initial
data.
In this framework the simplest non-trivial correlator is the second order one,
\begin{equation}
 \langle  \hat \psi_i^\ast \hat \psi_j \rangle = \langle  |\hat \psi_i| |\hat \psi_j| e^{\imath \left(\theta_i-\theta_j \right)} \rangle = n_i \delta (\mathbf{k}_i-\mathbf{k}_j),
\end{equation}
where the quantity $ n_i \equiv n(\mathbf{k}_i,t) $ is called {\it wave-action} spectrum.
The RPA assumption allows one to close the system by the Wick decomposition,
a mechanism which splits the higher order correlators as sums of products of second order correlators
\cite{cln0,cln1,cln2}.
This procedure, when applied to the GPE model, leads to the following four-wave {\it kinetic equation}
for the evolution of the wave-action spectrum \cite{dyachenko:1992hc},
\begin{eqnarray}\label{kin4}
\frac{\partial n_1}{\partial t} & = &4 \pi \int n_1 n_2 n_3 n_ 4 \left( \frac{1}{n_1} + \frac{1}{n_2} - \frac{1}{n_3} - \frac{1}{n_4} \right) \nonumber \\
& & \times \delta(\mathbf{k}_1+\mathbf{k}_2-\mathbf{k}_3-\mathbf{k}_4) \delta(\omega_1+\omega_2-\omega_3-\omega_4) d\mathbf{k}_{234} , \label{kineticequation}
\end{eqnarray}
where $d\mathbf{k}_{234} \equiv d\mathbf{k}_{2} d\mathbf{k}_{3} d\mathbf{k}_{4}$.

This is an integro-differential equation which models ``wave collisions'', in analogy with Boltzmann kinetic equation for the
particle collisions.
Physically, it says that for times much longer that the  fast wave periods   $ T_i = 2\pi/\omega_i $, the wave amplitudes
 $ \tilde{\psi}_i $ are effectively coupled  only if they satisfy the following resonant conditions,
\begin{equation}
\begin{split}
&\mathbf{k}_1+\mathbf{k}_2=\mathbf{k}_3+\mathbf{k}_4 \\
&\omega_1+\omega_2=\omega_3+\omega_4.
\end{split}
\end{equation}
Equation  (\ref{kineticequation})  has the following  invariants, $ M=\int n(\mathbf{k}, t) d\mathbf {k} $, $ {\bf P}=\int \mathbf{k} n(\mathbf{k}, t)d\mathbf{k} $, and $ E=\int \omega(\mathbf{k}) n(\mathbf{k}, t) d\mathbf{k} $ that are respectively the total number of particles (or the mass), the momentum and the energy. Moreover, the dynamics described by the  kinetic equation is irreversible in time and an entropy measure can be defined.
Three trivial functions that cancel the integrand in  (\ref{kineticequation}) are $ n_k^{(1)}= A $, $ n_k^{(2)}= (\mathbf{B} \cdot \mathbf{k})^{-1} $ and $ n_k^{(3)}=C k^{-2} $ with $ A, \mathbf{B},  $ and $ C $ constants. By combining these solutions, the general thermodynamic solution takes the form:
\begin{equation}
\label{eq:RJ}
n_k^{(RJ)} = \frac{T}{\mu +(\mathbf{u}\cdot\mathbf{k})+ k^2},
\end{equation}
where $ \mu $ is a  chemical potential constant, $ \mathbf{u} $ is a constant macroscopic velocity and $ T $ is a temperature of the system. In an isotropic field, the macroscopic velocity is zero and the relation \eqref{eq:RJ} assumes the form  known in literature as Rayleigh-Jeans (RJ)
thermodynamic equilibrium distribution. As the distribution is not convergent for large $ k $'s, the temperature and the chemical potential can be evaluated based on the known  mass and energy of the system only by introducing an ultraviolet cutoff, as proposed in \cite{PhysRevLett.95.263901}.

 In presence of an external forcing and damping, besides the  RJ distribution, two other steady   solutions may exist, namely the KZ spectra discussed in the introduction.
   These solutions
   have the form of a power-law  $ n_k = c k^{-\alpha} $, where
    $ c $ is a dimensional constant. They can be obtained analytically by applying a change of integration variables known as  the Zakharov transformation \cite{zakharov41kst}, or by a dimensional analysis
\cite{connaughton2003dimensional}.
The value of the exponent $ \alpha $ depends on the particular wave model. KZ solutions correspond to constant fluxes of positive
conserved quantities in
 the scale space (turbulent cascades).

The GPE model has two positive invariants, the mass $M$ and the energy $E$.
It can be shown (see \ref{ap:fluxes}) that energy has a direct cascade,
i.e. from large to small scales, while the mass cascade is inverse, from small to large scales.
The corresponding KZ exponents are respectively $ \alpha_E=3 $ and $ \alpha_M=7/3 $.
Hereafter in this work, we present results in terms of the one-dimensional spectrum  $ n_{1D}(k) $, obtained from $ n(k) $
by integrating out the angular variables, i.e.  $ n_{1D}(k) = 4 \pi k^2 n_k $.
In this notation the KZ solutions are
\begin{equation}
\begin{split}
&n_{1D}^{(E)}(k)= 4\pi c^{(E)} k^{-1} \\
&n_{1D}^{(M)}(k)=4\pi c^{(M)} k^{-1/3}.
\end{split}
\label{eq:KZ}
\end{equation}
These solutions represent respectively constant fluxes of $M$ and $E$ and, therefore, they are sustained by an external  forcing and a dissipation. Even tough the KZ spectra are found for infinite inertial ranges,
 they are expected
   in finite systems provided the scales of $ \mathcal{F} $ and $ \mathcal{D} $ are widely separated in Fourier space and
   provided the interactions of the wave modes are local.
   The locality of the KZ solutions (\ref{eq:KZ}) is checked  in  \ref{ap:locality}: the energy cascade turns out to be marginally nonlocal and its locality is restored by  a logarithmic correction, while the  inverse particle cascade is local.

\subsection{Transition to three-wave interactions}
Equation (\ref{kin4}) describes a four-wave interaction process which is responsible for the direct and the inverse cascades. When the inverse cascade is not damped at low $k$'s, it  leads to accumulation of  particles at these scales which can alter the four-wave dynamics.
Respectively, the {\it zero-mode} $ n_{\mathbf{k}=0}(t) $ (related to the uniform part of the field $ \psi $ in physical space) will grow, which
can be interpreted as a Bose-Einstein condensation process.
When the condensate fraction becomes large,  the kinetic equation   (\ref{kin4})  ceases to be valid.
Suppose that a large fraction of the wave-action is present at the zero-mode,
thereby $ \psi(\mathbf{x}, t)=c(t) + \epsilon\phi(\mathbf{x}, t) $, where $ \phi $ represents
small fluctuations ($ \epsilon \ll 1 $). By substituting this ansatz into GPE (\ref{nls0}), we find,
at the order $ \epsilon^0 $, the evolution equation for the {\it condensate fraction},
\begin{equation}
\imath \frac{\partial c}{\partial t} - |c|^2 c=0.
\end{equation}
Its  solution is $ c(t)=c_0 e^{-\imath c_0^2 t} $, where $c_0$ is a real positive constant.
Thus, the condensate amplitude  rotates in the complex plane with an
 angular velocity proportional to its square modulus.
In the next order in $\epsilon$, we obtain a linear equation for the fluctuations on  the condensate background:
\begin{equation}
\imath \frac{\partial \phi}{\partial t}(\mathbf{x}, t) + \nabla^2 \phi(\mathbf{x}, t) - 2c_0^2 \phi(\mathbf{x}, t) + c_0 e^{-2\imath c_0^2 t} \phi^\ast (\mathbf{x}, t)=0.
\label{phi}
\end{equation}
Diagonalizing the linear dynamics in Fourier space, one can show that the linear wave modes
oscillate at  the Bogoliubov frequency \cite{dyachenko:1992hc,zakharov2005dbe},
\begin{equation}
\omega(\mathbf{k})=\pm {k} \sqrt{{k}^2+2c_0^2}.
\label{eq:bogoliubov}
\end{equation}
In the limit of small $ k $'s or  strong condensate fraction $c_0$, when $k^2 \ll c_0^2$,
the fluctuations are acoustic waves with the speed of sound $ \omega/k \approx \sqrt{2c_0^2} $.

In the next order in $\epsilon$, when  weak nonlinearity is taken into account for the fluctuations,
 it is possible to use the WWT theory and derive a kinetic equation describing  three-wave interactions of the Bogoliubov sound
  \cite{dyachenko:1992hc,zakharov2005dbe}:
\begin{equation}
\frac{\partial n_1}{\partial t} = \int (R_{231} - R_{123} - R_{312}) d\mathbf{k}_{12},
\label{eq:3wave}
\end{equation}
where
\begin{equation}
R_{123} = 2\pi |V_{123}|^2 \delta(\mathbf{k}_1-\mathbf{k}_2-\mathbf{k}_3) \delta(\omega_1-\omega_2-\omega_3) (n_2 n_3 - n_1n_2 - n_1 n_3)
\end{equation}
and the analytical form of the scattering matrix $V_{123}$ is given in \cite{zakharov2005dbe}.
Equation (\ref{eq:3wave}) describes  three-wave interactions $ 1 \leftrightarrows 2 $  which conserve only the energy
 and not the mass. Thus, only the energy cascade KZ solution is relevant to this regime.
For the large-scale (strong condensate) limit $k^2 \ll c_0^2$, the
 direct cascade KZ spectrum takes the form $ E_{1D}(k) \sim k^{-3/2} $ predicted by Zakahrov and Sagdeev for the
  3D acoustic WWT \cite{zakharov1970sat}. Because most of the wave-action in this case is in the condensate, the 1D energy spectrum is $ E_{1D}^{(E)}(k) \sim c_0^2 n_{1D}^{(E)}(k) $ \cite{zakharov2005dbe}. Therefore, the wave-action spectrum for the energy cascade in the acoustic
  regime is
\begin{equation}
n_{1D}^{(E)}(k) \sim k^{-3/2}.
\end{equation}

\subsection{Critical balance conjecture}
In some physical situations, wave turbulence fails to be weak, and the wave spectrum saturates at a critical shape such that the linear term is of the same size as the nonlinear term for each mode ${\bf k}$. Such a \em critical balance \em appears to be typical for a wide range of
physical systems, ranging from Magneto-Hydrodynamic turbulence \cite{goldreich1995tti}, to the rotating and stratified geophysical systems \cite{nazarenko-cb-paper}.
The most famous example is the Phillips spectrum of the gravity water waves \cite{phillips:367511, korotkevich:074504}, in which case the saturation at the critical value occurs due to wave breaking.

In the GPE model, similar situation may occur when an equivalent of wave breaking process is active in the system. Namely, we will see that when the low-${\bf k}$ range is over-dissipated by strong hypo-viscosity ({\bf RUN 3}), the inverse particle cascade tends to accumulate at low ${\bf k}$'s ({\it infrared bottleneck}) until a critical balance is reached and the spectrum is saturated.
Indeed, for the inverse cascade to exist the wave turbulence must be weak, because only then the Fj{\o}rtoft argument works (see \ref{fjortoft}). However, when the linear and the nonlinear terms, locally in Fourier space, are of the same order, the inverse cascade stops and so does
the infrared bottleneck  accumulation. This is precisely the mechanism of reaching the critical balance condition in this case.

Now we present an estimate for the critical balance spectrum in the GPE model.
Equating the linear and the nonlinear terms in Fourier space gives
\begin{equation}
\begin{array}{rcl}
k^2 |\tilde{\psi}_k|   & \sim & |\tilde{\psi}_k|^3 k^6 \\
\Rightarrow \; k^{-4} & \sim & |\tilde{\psi}_k|^2.
\end{array}
\end{equation}
Note that we have replaced each $ d\mathbf{k} $ integration  by $ k^3 $ thereby assuming that only the wave amplitudes with similar $k$'s
 are correlated in Fourier space.
Thus, for the 1D wave-action spectrum  in the critical balance regime, we have
\begin{equation}
n_{1D}^{(CB)}(k) \sim k^{-2}.
\end{equation}

\section{The numerical experiments}\label{numerics}
As we want to understand the basic properties of the GPE turbulence,
 we choose to deal with the simplest configuration: triple periodic boundary condition and uniform mesh grid. With this choice we can use the discrete Fourier transforms which are numerically fast \cite{press:1992rm, url:fftw}. In our numerics the complex wave field $ \psi $ is a double precision variable defined in space over $ 256^3 $ points. The simulation box has side $ L=256 $ and so the Fourier space width is $ L_k = 2\pi $ with resolution $ \Delta k=2\pi/256 $. Without considering for the moment the forcing and the dissipation
  terms, the GPE model (\ref{nls0}) can be written in the physical space as a sum of a linear operator $ \mathcal{L} $ and a nonlinear one $ \mathcal{G} $
\begin{equation}
\imath \frac{\partial \psi}{\partial t} = \mathcal{L} \psi + \mathcal{G} \psi,
\end{equation}
where $ \mathcal{L} = -\nabla^2 $ and $ \mathcal{G} = |\psi|^2 $. We then use a split step method to solve separately the contributions
 of the two operators in time. This choice is very useful because the linear operator has the exact solution in Fourier space, $ \tilde{\psi}(\mathbf{k}, t+\Delta t)=\tilde{\psi}(\mathbf{k}, t) e^{-\imath |\mathbf{k}|^2 \Delta t} $, while the nonlinear one, due to the conservation of the mass in the system, has the analytic solution $ \psi(\mathbf{x}, t+\Delta t)=\psi(\mathbf{x}, t) e^{-i |\psi(\mathbf{x}, t)|^2 \Delta t} $ in the physical space. At each time step $ \Delta t $, we first evaluate the linear part in Fourier space and then use this temporary solution to solve the nonlinear part in the physical space. With this choice, the numerical error in the algorithm is only due to the time-splitting \cite{bao:2006zk}. The time step is always $ \Delta t =0.5 $, chosen to be of the order of the shortest linear time $ 2\pi/\omega_{max} $.

Concerning the forcing and the dissipation, it appears
 convenient to control $ \mathcal{F} $ and $ \mathcal{D} $ directly in Fourier space to gain a wide inertial range.
  The forcing term acts to inject mass and energy in the system. In all simulations, $ \mathcal{F} $ modifies the first (linear)
  calculation half-step  as  follows,
\begin{equation}
\tilde{\psi}(\mathbf{k}, t+\Delta t)= \left\{
\begin{array}{ll}
\tilde{\psi}(\mathbf{k}, t) e^{-\imath |\mathbf{k}|^2 \Delta t} + A f_0 e^{\imath \varphi(\mathbf{k}, t)}, \ \ & k_{\min} \le |\mathbf{k}| \le k_{\max} \\
\tilde{\psi}(\mathbf{k}, t) e^{-\imath |\mathbf{k}|^2 \Delta t}, & |\mathbf{k}|<k_{\min} \cup |\mathbf{k}|>k_{\max},
\end{array}
\right.
\label{eq:forcing}
\end{equation}
where $ A \simeq 1.62 \times 10^{-3} $. Thus the pumping add mass and energy in the ring region $ k_{\min} \le |\mathbf{k}| \le k_{\max} $ with the function $ \varphi(\mathbf{k}, t) $ uniformly distributed in $ [0, 2\pi) $ and statistically independent at each $\bf  k $-space and
each time step. We choose the forcing  at relatively large scales, with $ k_{\min}=9 \Delta k $ and $ k_{\max}=10 \Delta k $.
A dissipation at high wave numbers is added to halt the direct cascade and to prevent termalization effects.
We find that an hyper-viscous term $ \mathcal{D}_h = \imath\nu_h(\nabla^2)^8 \psi(\mathbf{x}, t) $, where $ \nu_h=2 \times 10^{-6} $,
is effective in absorbing the high-$k$ spectrum and in preventing the aliasing and the bottleneck effects.
This term is also added to the linear operator to evaluate it efficiently in Fourier space.
Finally, different types of dissipations at the large scales can be chosen.
In this manuscript we report three different setups, briefly summarized in Table \ref{tab:cases},
whose results are discussed in the following.
\begin{table}
\begin{center}
\begin{tabular}{| c | c | c | c |}
\hline
cases & forcing & dissipation at low $ k $'s & dissipation at high $ k $'s \\
\hline
{\bf RUN 1} & $ f_0=0.1 $ & none &  $ \mathcal{D}_{h}=\imath \nu_h (\nabla^2)^8 \psi(\mathbf{x}, t) $ \\
{\bf RUN 2} & $ f_0=0.1 $ & $ \hat{\mathcal{D}}_{l}=\imath \theta(k^\star-|\mathbf{k}|)\tilde{\psi}(\mathbf{k}, t) $ &  $ \mathcal{D}_{h}=\imath \nu_h (\nabla^2)^8 \psi(\mathbf{x}, t) $ \\
{\bf RUN 3} &  various & $ \mathcal{D}_{l}=\imath \nu_l (\nabla^{-2})^8 \psi(\mathbf{x}, t) $ &  $ \mathcal{D}_{h}=\imath \nu_h (\nabla^2)^8 \psi(\mathbf{x}, t) $ \\
\hline
\end{tabular}
\end{center}
\caption{Summary of the different numerical simulations performed. Here the hat operator $ \hat{\cdot} $ stands for the Fourier-space
 operator, $ \nu_h = 2 \times 10^{-6} $, $ \nu_l =1 \times 10^{-18} $, $ \mu = 1 \times 10{-4} $, $ k^\star=9 \Delta k $ and the forcing is defined in (\ref{eq:forcing}).}
\label{tab:cases}
\end{table}

\subsection{{\bf RUN 1} - free condensate growth}
\begin{figure}
\includegraphics[scale=1.6]{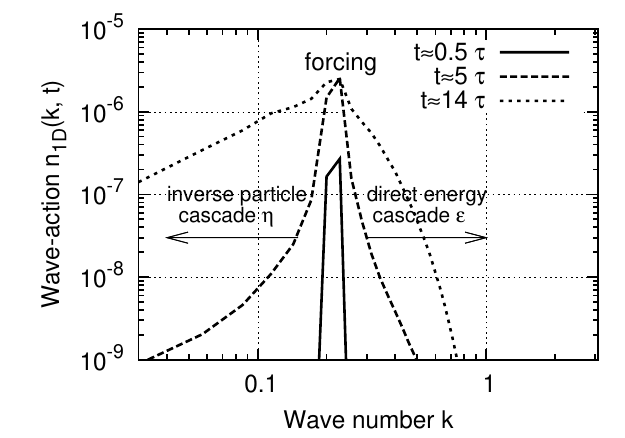}
\caption{Wave-action spectrum $ n_{1D}(k, t) $ during initial stages of simulation. The time unit $ \tau $ refers to the linear time of the forcing scale.\label{in-spectra}}
\end{figure}

\begin{figure}
\includegraphics[scale=1.6]{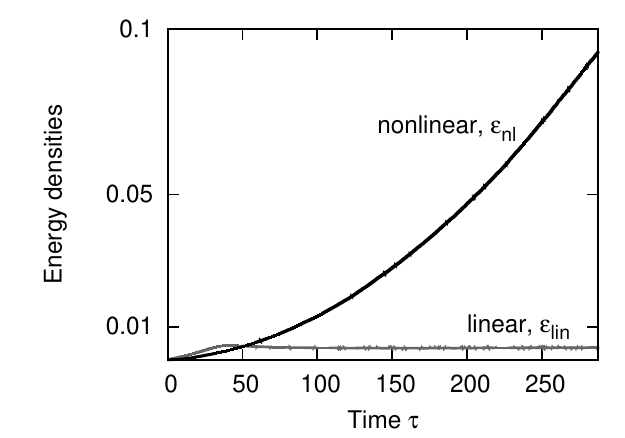}
\caption{Linear $ \mathcal{E}_{lin}(t) $  and nonlinear $ \mathcal{E}_{nl}(t) $ energy densities during the {\bf RUN 1}. \label{hv-energy}}
\end{figure}

Here, we study the evolution of the GPE system, initially empty, without any dissipation at low $ \mathbf{k} $'s.
At every time step, the forcing term inputs  mass $ \Delta M(t) $ and energy $ \Delta E(t) $, so that $M$ and $E$
 start to grow. Even if the forcing is acting at a particular wave-number range,
  the nonlinear interactions cause mass and energy  transfers in Fourier space.
  We have chosen the forcing coefficient $ f_0 $ such that the transfers become efficient for the
  time step considered, namely it chosen to be sufficiently  big to prevent sandpile effects characteristic to
  mesoscopic turbulence \cite{nazarenko:2006ta,zakharov:2005jt}.
   One can see in Fig. \ref{in-spectra} that at early stages of the simulation the wave-action
   spectrum $ n_{1D}(k, t) $ evolves and spreads over the wave-numbers space:
   the energy and the particles undergo a direct and an inverse cascades respectively.
After an initial transient time, the linear energy density $ \mathcal{E}_{lin}(t) $ stops to grow, as clearly visible in Fig. \ref{hv-energy}. Linear energy weights at the high wave number part of the spectrum and this saturation is an evidence that its
transfer to the
 small scales is now absorbed by the hyper-viscosity. On the contrary, as no dissipation is present at the large scales,
  the inverse particle cascade is not arrested and the  nonlinear energy density $ \mathcal{E}_{nl}(t) $ continues to increase.

\subsubsection{Spectra}
\begin{figure}
\includegraphics[scale=1.6]{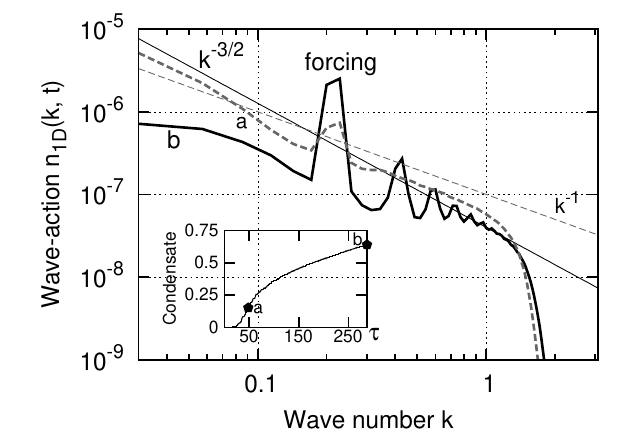}
\caption{Wave-action spectrum $ n_{1D}(k, t) $ at two different stages of {\bf RUN 1}.
 The four-wave and the three-wave WWT
 predictions are also indicated by the lines with slopes $-1$ and $-3/2$ respectively.
  Inset: evolution of the condensate component $ c_0(t) $.\label{spectrum-hv}}
\end{figure}

\begin{figure}
\includegraphics[scale=1.6]{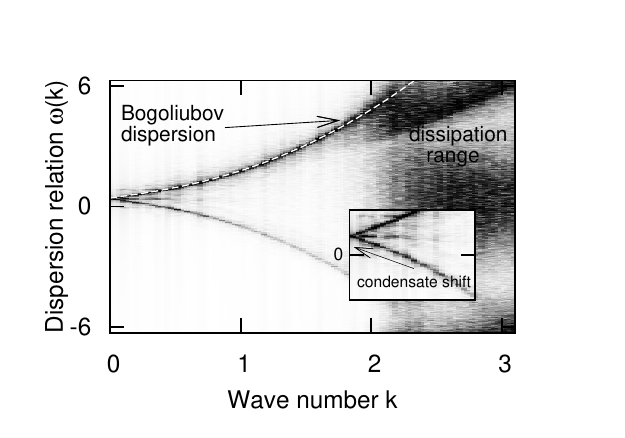}
\caption{Dispersion relation at the final stage of {\bf RUN 1}, when the condensate fraction is strong. The white dashed line is the Bogoliubov dispersion (only the upper branch). Inset: a zoom on the small $ k $'s zone to appreciate the shift due to the presence of the condensate.\label{dispersion-hv}}
\end{figure}

In Fig. \ref{spectrum-hv} we present the 1D
wave-action spectrum at two different stages after the stabilization of the linear energy.
The spectrum plotted with dashed line is taken at early stages, when the linear and the nonlinear energy
densities are comparable: at this point there is a good agreement with the WWT $ k^{-1} $ prediction, which is also plotted.
The condensate component $ c_0(t) $ continues to increase during the simulation, as shown in inset to Fig. \ref{spectrum-hv},
 and  a well defined series of peaks appears in the wave-action spectrum at the final stage (continuous line).
 Such behavior is a clear sign  of three-wave interactions, as reported in \cite{falkovich1988efc}.
Note that the late stage spectrum is consistent with the three-wave Zakharov-Sagdeev WWT prediction $ k^{-3/2} $.
The acoustic three-wave regime is also confirmed by evaluating the dispersion relation by measuring the wavenumber-frequency Fourier transform. 
The result, presented in Fig. \ref{dispersion-hv}, shows the presence of two branches (one for the Bogoliubov mode and another for its conjugate) shifted by the condensate velocity oscillation $ c_0^2 $, in excellent agreement with the corresponding Bogoliubov dispersion (\ref{eq:bogoliubov}).

The observed evolution of the spectrum can be summarized  as follows. (a) At initial times  the condensate wave amplitude $ \tilde{\psi}_{\mathbf{k}=0} (t) $ is of the same order as the amplitude of other modes and the dynamics is well described by the four-wave WWT regime. (b) As the inverse cascade is not halted, mass accumulates at low $ \mathbf{k} $'s and strong turbulence takes place in this transition. (c) Finally when the zero-mode becomes dominant over the fluctuations the wave turbulence is again weak and well described by the three-wave WWT.

\subsubsection{Condensation and density PDF}
 
\begin{figure}
\includegraphics[scale=1.6]{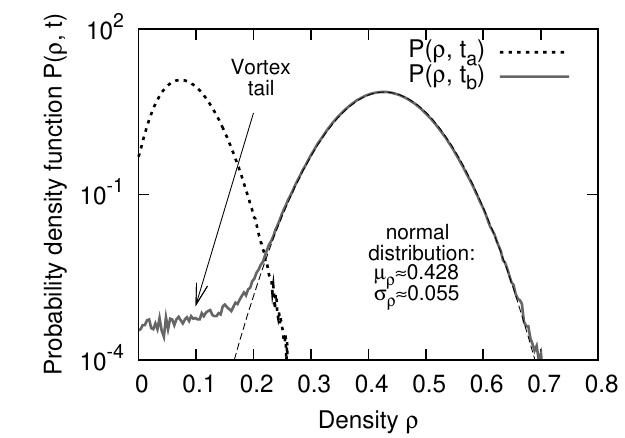}
\caption{Probability density function of the density field $ \rho $ at different time: early stages ($ t_a $) and final stage ($ t_b $) in {\bf RUN 1}. A normal distribution with mean $ \mu $ and standard deviation $ \sigma $ are indicated.\label{fig:1_desity_pdfs}}
\end{figure}

 To access further information about  the statistics of the transition from the four-wave to the three-wave regime during the condensation process, in Fig. \ref{fig:1_desity_pdfs} we show the probability density function (PDF) of the density field $ \rho $ at initial and final stages.
At early stages the density remains small with a lot of low density regions present, which indicates presence of numerous ``ghost'' (weakly nonlinear)  vortices. In fact, in the ideal four-wave WWT dynamics, the wave field would be nearly Gaussian, and the respective density $\rho$ would have an exponential (Rayleigh) PDF, which would be a straight line (with a negative slope) in Fig. \ref{fig:1_desity_pdfs}. We see a significant deviation from the Rayleigh behavior, which means that even at an early stage (at time $t_a$ when the four-wave KZ spectrum is reported) the statistics already differed from the Gaussian.
The development of a maximum on the PDF of $\rho$ is a signature of the emerging condensate.
 As the inverse cascade is not halted the condensate density keeps growing in time. At final stages (at time $t_b$ when the three-wave KZ spectrum is reported) the density field shows a normal distribution behavior in the core of the PDF with  a tail remaining at low density regions corresponding to vortices (which are now strongly nonlinear).
  Note that the mean value of $\rho$ at this time (0.428) is much greater than the standard deviation (0.055) which means that the condensate
    density is much stronger that the  Boboliubov fluctuations about the mean density. This is a clear sign of the three-wave WWT.

\subsubsection{Vortices and velocity PDF}

\begin{figure}
\includegraphics[scale=0.4]{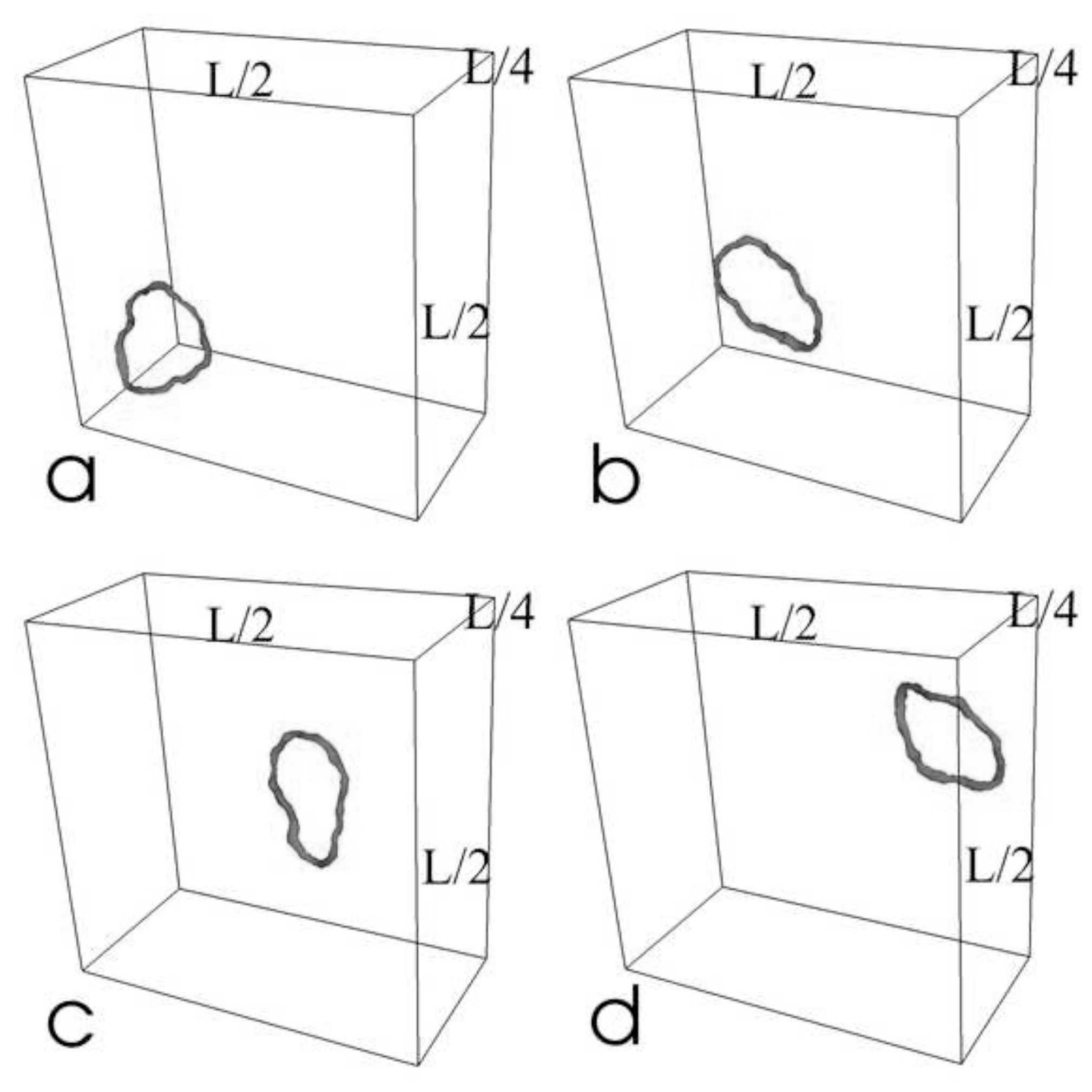}
\caption{Snapshots of iso-surfaces of low density region with threshold $ \rho_{thr}=0.1 $ (the mean density is $ \langle \rho \rangle \simeq 0.43 $). The box is $ 1/2\times 1/2 \times 1/4 $ the computational domain. The frames are taken approximately every $ 3 \tau $.\label{fig:1_vortex}}
\end{figure}

     Now let us focus on the vortex component represented by the low-$\rho$ PDF tail at $t=t_b$. By plotting the iso-surfaces with small density threshold ($ \rho_{thr}=0.1 $) only one quantum vortex is found in the computational box. We plot in Fig. \ref{fig:1_vortex} its evolution, showing only a part of the total box. The vortex has a ring shape and it propagates in the direction of the ring axis. The  vortex core radius is consistent with the healing length estimate $ \xi \simeq 1.5 = 1.5 \Delta x $.
  Propagating Kelvin waves can be observed on the vortex line. We will argue below that these waves are crucial for understanding the $5/3$-spectrum of the incompressible kinetic energy.

\begin{figure}
\includegraphics[scale=1.6]{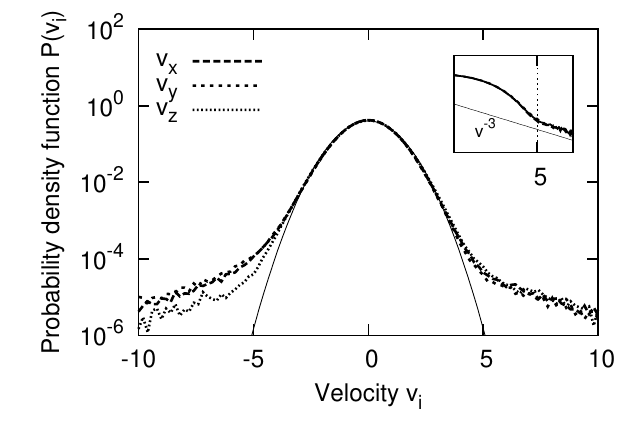}
\caption{Probability density function of the single velocity components at the final stage in {\bf RUN 1}. Inset: zoom on  
  the positive tail of the distribution in log-log coordinates.\label{fig:1_pdf_vel}}
\end{figure}

In Fig. \ref{fig:1_pdf_vel} we plot the PDF of the single velocity  components (the data are normalized in order to compare
 different distributions). 
The velocity PDF appears to have a dominant Gaussian core, which is consistent with the three-wave WWT.
In addition, the velocity PDF power-law  tail  with exponent
 $ -3 $ is another signature of the thin vortex lines, whose velocity field falls off inversely proportional to the distance
  from the line at short distances. Such power-law velocity PDF's where observed
 in superfluid turbulence  experimentally  \cite{paoletti:154501} and numerically \cite{PhysRevLett.104.075301} and interpreted as 
   an evidence of quantum vortices.

\subsubsection{Incompressible energy spectrum}

\begin{figure}
\includegraphics[scale=1.6]{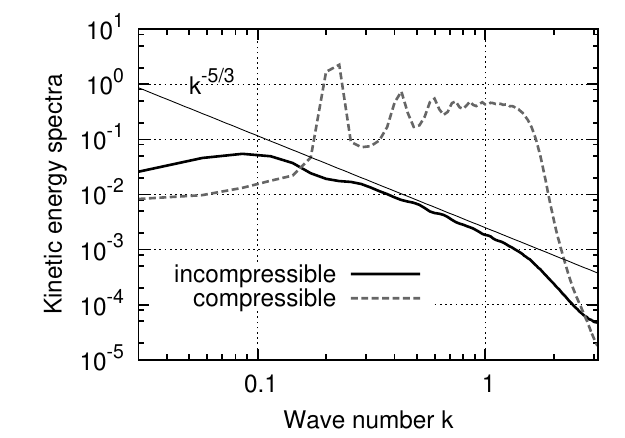}
\caption{Compressible and incompressible kinetic energy spectra measured at final stage of simulation in {\bf RUN 1}. A $ k^{-5/3} $ showin classical Kolmogorov prediction is also plotted.\label{fig:1_kin_spectra}}
\end{figure}

Spectra of the compressible and the incompressible kinetic energy at late time ($t=t_b$) are presented in Fig. \ref{fig:1_kin_spectra}.
The compressible part is dominant and has the same features as the ones already discussed for the wave-action spectrum: it shows  the peaks 
(harmonics of the forcing scale) and it follows the  Zakharov-Sagdeev spectrum (with exponent $-3/2+2 =1/2$).
 The incompressible spectrum formally coincides with the classical Kolmogorov $ 5/3 $-law.
  But we have seen in Fig. \ref{fig:1_vortex} that only one quantized vortex ring remains in the system at this time, so it is impossible
  for the Kolmogorov theory, developed for the continuous classical vorticity fields, to be relevant in this case.
  Resolution to this paradox is suggested in \cite{lvov-et-al-5/3}. In short, the $ -5/3 $ slope is produced
  by an energy cascade carried by Kelvin waves (seen in
  Fig. \ref{fig:1_vortex}).
  It turns out \cite{LN-spectrum} 
  that the Kelvin wave energy spectrum also has exponent $ -5/3 $ and the coincidence with the Kolmogorov
  exponent is purely coincidental (the Kelvin-wave and the Kolmogorov spectra have different pre-factors). 



\subsection{{\bf RUN 2} - friction at large scales}
\begin{figure}
\includegraphics[scale=1.6]{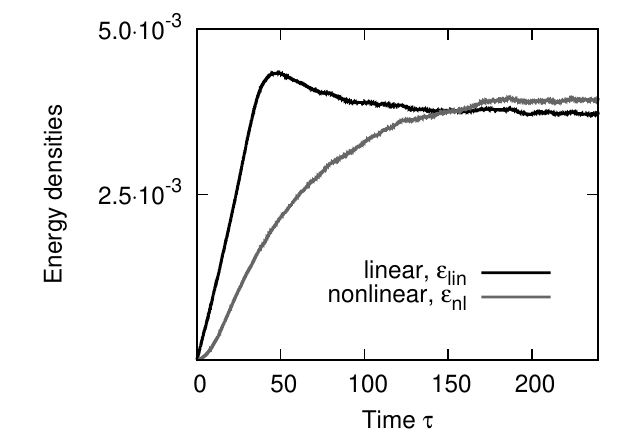}
\caption{Linear $ \mathcal{E}_{lin}(t) $ and nonlinear $ \mathcal{E}_{nl}(t) $ energy density evolution in the presence of large scale friction ({\bf RUN 2}). \label{hvf-energy}}
\end{figure}

\begin{figure}
\includegraphics[scale=1.6]{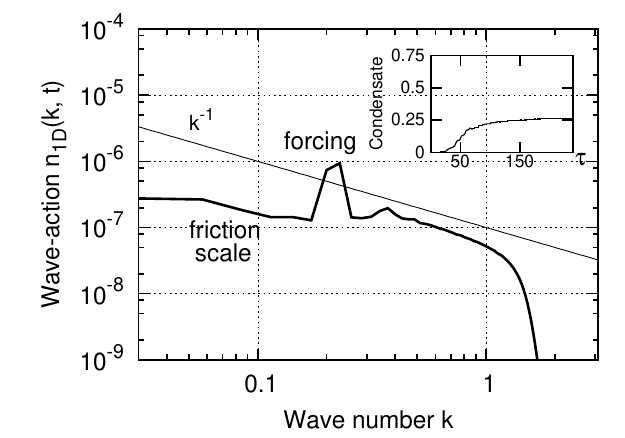}
\caption{Wave-action spectrum $ n_{1D}(k, t) $ at final stage of simulation in the presence of friction. The WWT four-wave prediction $ k^{-1} $ is 
shown by the straight line. The inset shows the condensate evolution $ c_0(t) $. \label{spectrum-hvf}}
\end{figure}

In the previous run no steady turbulent state has been reached because of the presence of the inverse cascade. 
To stay in the four-wave weak turbulence regime  and to 
avoid the condensate growth, an effective friction term at large scales will now be added. 
This term is written directly in Fourier space and results in
\begin{equation}
\hat{\mathcal{D}}_l = \imath \mu \theta(k^\star-|\mathbf{k}|) \tilde{\psi},
\end{equation}
where $ \theta $ is the Heaviside step function, $ k^\star = 9 \Delta k $ is the lowest forced wave-number and $ \mu=1 \times 10^{-4} $ is a friction coefficient. This coefficient is optimally chosen to stop the inverse cascade without altering the direct energy cascade. With this choice both the nonlinear and the linear energy densities reach a constant value during the simulation, as plotted in Fig. \ref{hvf-energy}.
The final stage wave-action spectrum, presented in Fig. \ref{spectrum-hvf}, agrees with the $ k^{-1} $ four-wave WWT prediction. The condensate growth, shown in the inset, is halted by the friction. Agreement with the four-wave WWT may seem surprising because, according to 
 Fig. \ref{hvf-energy}, the nonlinear energy exceeds the linear one. Our explanation is that the nonlinear energy  mostly
 resides in the condensate and forcing scales, whereas in the direct cascade range the modes are weakly nonlinear.

\begin{figure}
\includegraphics[scale=0.4]{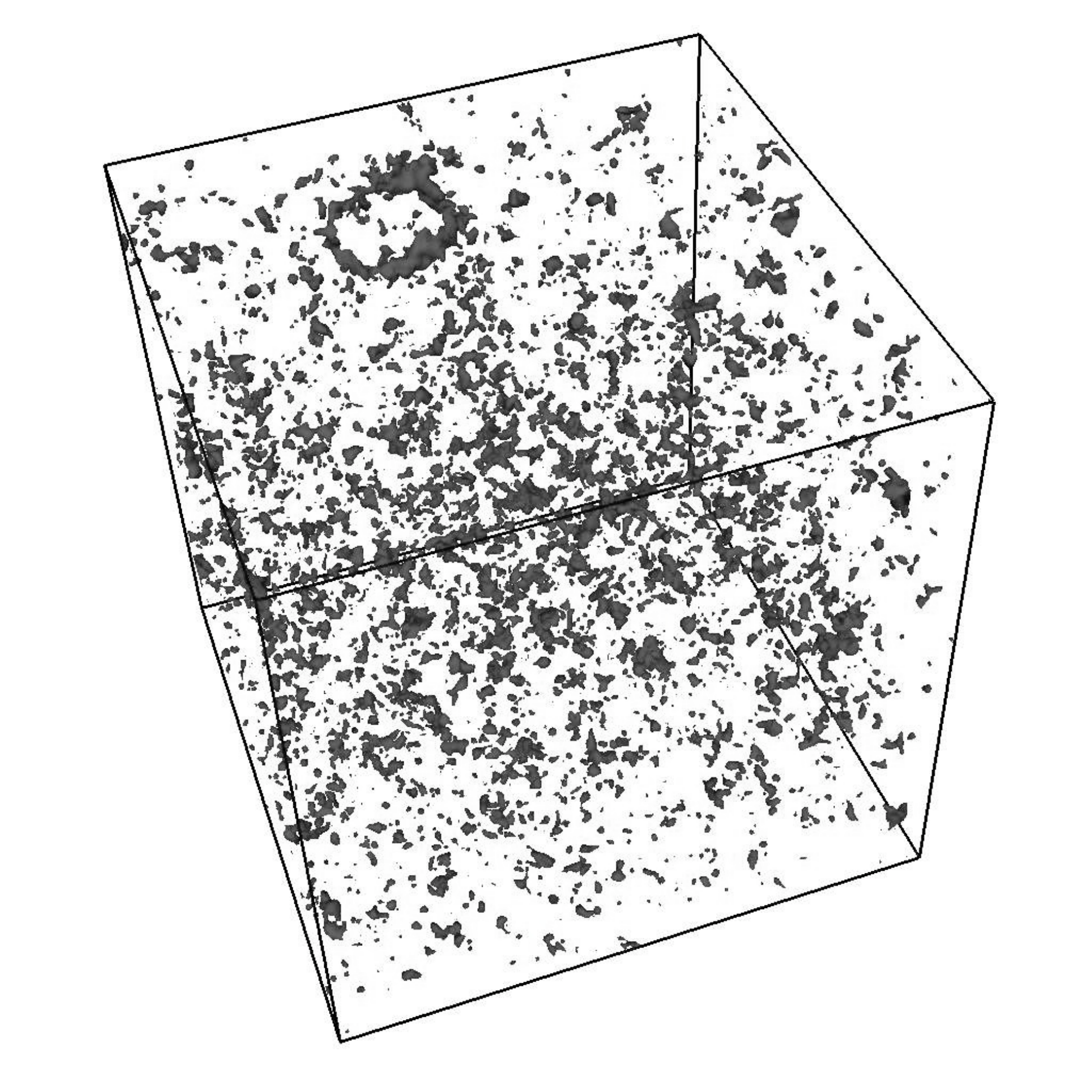}
\caption{Low density region of the density field, threshold $ \rho_{thr}=0.015 $, in presence of friction ({\bf RUN 2}).\label{density-fric}}
\end{figure}

We now turn our attention to the statistically 
steady state distributions in the physical space. The PDF of the density field $ \rho $, not showed here, 
looks similar to early stage distribution presented in Fig. \ref{fig:1_desity_pdfs}.
Here the average density is $ \langle \rho \rangle \simeq 8.31\cdot 10^{-2} $, which corresponds to healing length $ \xi \simeq 3.5 \Delta x $. The low density regions in the computational box are plotted in Fig. \ref{density-fric}.
A big fuzzy ring structure is present in the top of the figure. It is probably a single ring with large fluctuations, which create secondary small
vortex loops near the main vortex. 
Besides this ring, other uniform bubble-like low density regions are present in the box. These are small scale ghost vortices
 which are weakly nonlinear and short-lived (their typical  size is close to the resolution scale).
Thus we see that even though the wave field is mostly random, the coherent vortex structures are also seen in this 
regime.

\begin{figure}
\includegraphics[scale=1.6]{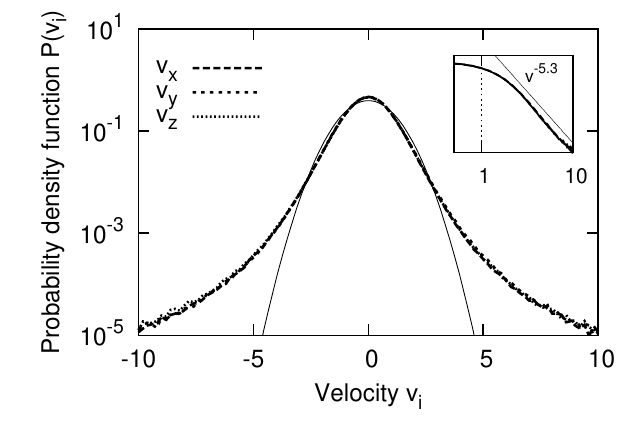}
\caption{Probability density function of  velocity components in the presence of friction ({\bf RUN 2}). Inset: zoom in log, log scale of the positive PDFs branch.\label{pdf_vel-fric}}
\end{figure}

\begin{figure}
\includegraphics[scale=1.6]{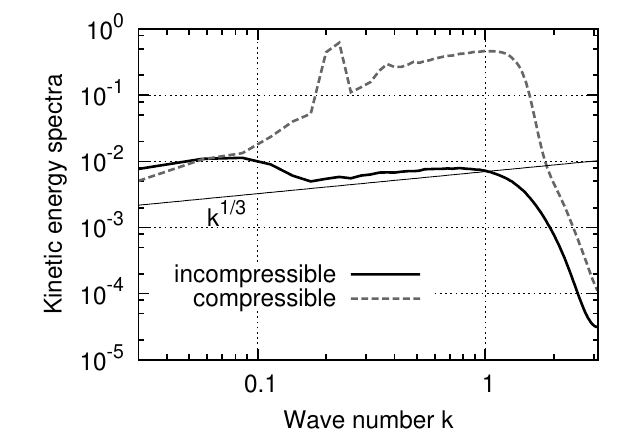}
\caption{Compressible and incompressible kinetic energy spectra in a steady state with friction ({\bf RUN 2}).\label{fig:2_kin_en}}
\end{figure}

The system in the fluid dynamics framework presents quite unexpected results, still remaining to be explained.
 The PDFs of the  velocity components are plotted in Fig. \ref{pdf_vel-fric}.
The PDF is isotropic and Gaussian in the core, but it has a power-law tail  $\sim v_i^{-5.3} $. 
We still do not have a theoretical explication for this exponent, although the power-law PDF tail with a different
exponent (-1) was previously predicted for WWT in \cite{cln0}. 
The compressible and the incompressible kinetic energy spectra are presented in Fig. \ref{fig:2_kin_en}.
The compressible spectrum does not show peaks (as in {\bf RUN 1}) indicating that this turbulence is not
acoustic. The incompressible spectrum shows
 a power-law behavior with exponent close to $ 1/3 $. Again, no theoretical explanation could be proposed.



\subsection{{\bf RUN 3} - hypo-viscosity dissipation}
\begin{figure}
\includegraphics[scale=1.6]{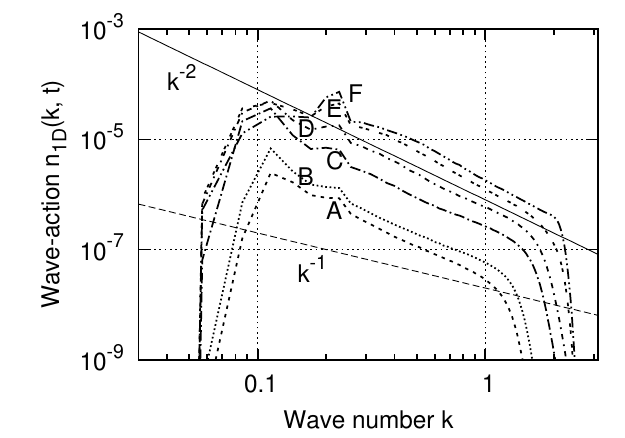}
\caption{Wave-action spectra $ n_{1D}(k, t) $ at the final stage of the simulation in the presence of  hypo-viscosity ({\bf RUN 3}) for different forcing coefficients: $ f_0=0.05 $ {\bf (A)}, $ f_0=0.1 $ {\bf (B)}, $ f_0=0.5 $ {\bf (C)}, $ f_0=1.0 $ {\bf (D)}, $ f_0=2 $ {\bf (E)}, $ f_0=3 $ {\bf (F)}. The $ k^{-1} $ WWT and $ k^{-2} $ CB predictions are also shown. \label{spectrum-all-hlv}}
\end{figure}

One can devise to stop the inverse cascade with a different type of dissipation at low $ k $'s.
For this, let us now choose a hypo-viscosity of the form
\begin{equation}
\mathcal{D}_l=\imath (\nabla^{-2})^8 \psi(\mathbf{x}, t).
\end{equation}
As this operator is singular in $ \mathbf{k}=0 $, we will separately remove, at each time step, the zero-mode in Fourier space. This choice still allows to reach steady turbulent states, but unexpectedly leads to different results with respect to {\bf RUN 2}. In Fig. \ref{spectrum-all-hlv} we present various steady state spectra evaluated with  different levels of the forcing $ f_0 $.
What emerges clearly  is that the spectra still have a power-law behavior in the inertial range, but
 it does not follow the WWT four-wave prediction $ k^{-1} $. Instead, we see the $ k^{-2} $ critical balance prediction
 for the wide range of the forcing amplitudes.

\begin{figure}
\includegraphics[scale=1.6]{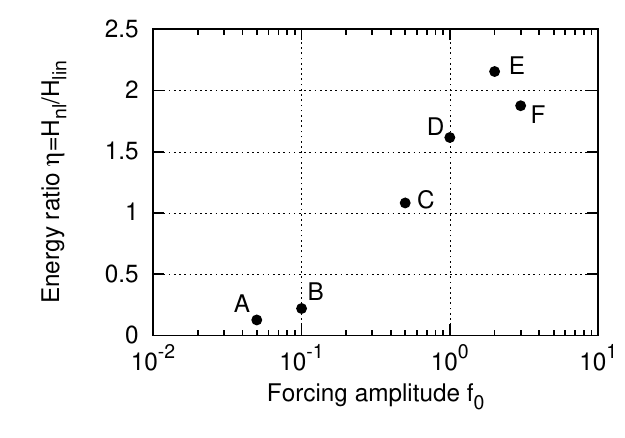}
\caption{Energy ratio $ \eta=H_{nl}/H_{lin} $ with respect to different forcing coefficients $ f_0 $ in the presence of hypo-viscosity ({\bf RUN 3}). For informations about labels see caption in Fig. \ref{spectrum-all-hlv}. \label{energy_ratio-hlv}}
\end{figure}

Our interpretation is the following. 
The hypo-viscosity causes an infrared bottleneck that alters the dynamics at the 
scales near the forcing: there the linear and the nonlinear energies become comparable. 
At this point the Fj{\o}rtoft argument can no longer apply  and the 
critical balance condition propagates into
 all the inertial range causing the observed $ k^{-2} $ spectra. This suggestion is 
 corroborated by the measurement of the ratio between nonlinear and linear energies $ \eta={H_{nl}}/{H_{lin}} $ 
 evaluated for various forcing coefficients $ f_0 $  illustrated in Fig. \ref{energy_ratio-hlv}.
 From these results it is clear that, for a wide range of the forcing coefficients (almost two orders of magnitude), 
 the energy ratio $ \eta $ remains always order one.

\begin{figure}
\includegraphics[scale=0.4]{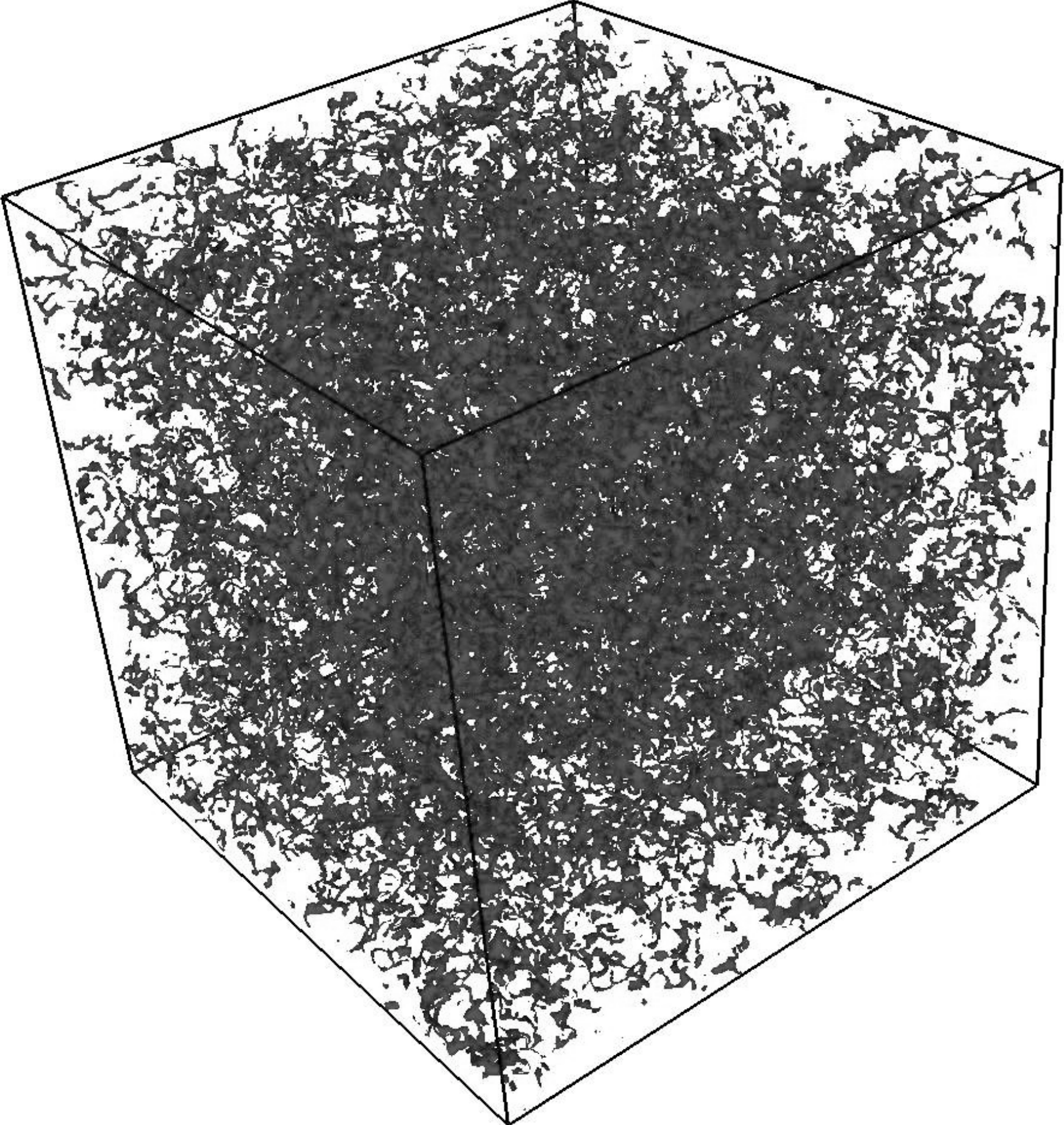}
\caption{Low density regions with threshold of $ \rho_{thr}=0.05 $ are plotted at the final stage with the presence of hypo-viscosity with $ f_0=1.0 $ ({\bf RUN 3}). \label{density-hypo}}
\end{figure}

As in the previous runs, we  look at the density field in the
 physical space to visualize vortices and other turbulent structures.
 For this we choose the case ({\bf D}) when $ f_0=1.0 $ (all other cases look similar).
  Again, the PDF of the density is similar to the early stage PDF in Fig. \ref{fig:1_desity_pdfs},
   and this is natural because there is no condensate in the present system. 
   The mean density is $ \langle \rho \rangle \simeq 4.31 \cdot 10^{-1} $ and so $ \xi \simeq 1.5 \Delta x $.
In Fig. \ref{density-hypo} we show the low density regions in the physical space
with a threshold $ \rho_{thr}=0.05 $. 
 This figure is qualitatively different from the previous ones (Fig.  \ref{fig:1_vortex} and Fig. \ref{density-fric}).
 Now very thin vortex structures fill completely the computational box and form a ``vortex tangle''.

\begin{figure}
\includegraphics[scale=1.6]{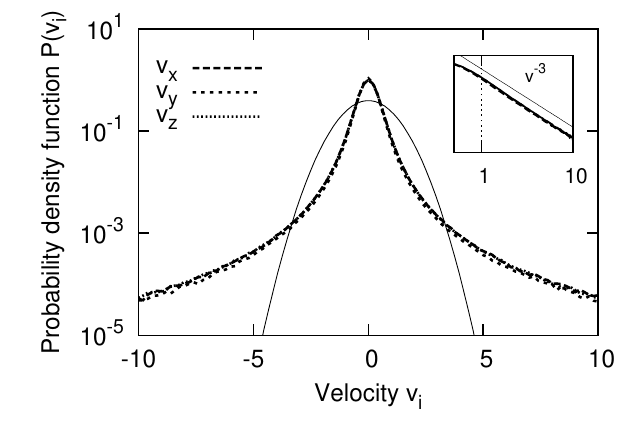}
\caption{PDF of single velocity components in the presence of hypo-viscosity with $ f_0=1.0 $ ({\bf RUN 3}). Inset: zoom  of the positive branch of the PDF in log-log coordinates.\label{pdf_vel-hypo}}
\end{figure}

\begin{figure}
\includegraphics[scale=1.6]{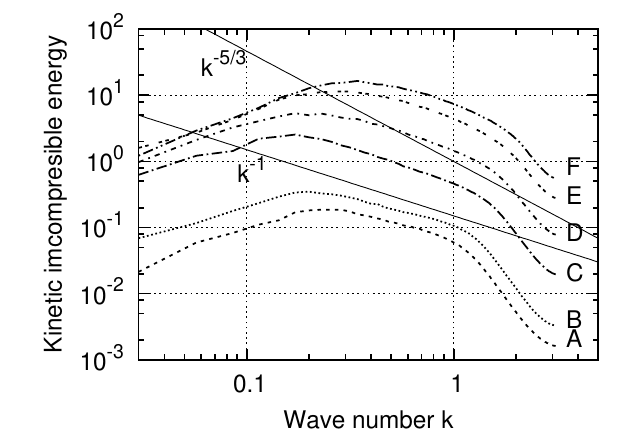}
\caption{Incompressible kinetic energy spectra for different forcing amplitude $ f_0 $ in presence of hypo-viscosity ({\bf RUN 3}). For information about the labels see caption in Fig. \ref{spectrum-all-hlv}.\label{fig:3_kin_en}}
\end{figure}

The one point PDF's of the single velocity components are plotted in Fig. \ref{pdf_vel-hypo}.
The PDF's, which show isotropy, are strongly non-Gaussian: the power-law behavior of $ v^{-3} $ is observed. 
This result  is  interesting because similar behavior was observed 
experimentally in \cite{paoletti:154501} and numerically in 
\cite{PhysRevLett.104.075301} and explained by presence of thin quantized vortex lines. 
We emphasize  the $ v^{-3} $ PDF behavior is dominant and not present just in the tail as in the {\bf RUN 1}.
This is because we have much more strong vortex lines in the {\bf RUN 3} than in the {\bf RUN 1}.
Indeed, as the condensate fraction is removed by this type of dissipation,
it is thus natural to think that the vortices fill the system at all scales and highly influence the velocity field. 
These vortex lines undergo frequent reconnections resulting in a sound emission. 
The incompressible kinetic energy spectra, for all the forcing amplitudes, are illustrated in Fig. \ref{fig:3_kin_en}.
No evidence of  $ -5/3 $ law is found in this (critical balanced) regime. 
All spectra have a power-law behavior in the inertial range with exponent near $ -1 $. A theoretical explanation of this exponent is
 still lacking.



\section{Conclusions}\label{conclusion}
In this paper, we have analyzed the turbulent states in the forced-dissipated 3D GPE model by  using the  direct numerical simulations. 
Introduction of the forcing and the damping is aimed at achieving statistically stationary turbulent cascades. 
We have focused our attention on the direct energy cascade by introducing a pumping term at relatively 
large scales and by  using hyper-viscosity  at small scales. We have studied three regimes with different 
dampings at the scales larger than the forcing scale: 
no dissipation ({\bf RUN 1}), a friction ({\bf RUN 2}) and a hypo-viscosity ({\bf RUN 3}).

{\bf RUN 1} is performed without any dissipation at large scales and so the inverse cascade, 
predicted by the WWT theory, causes condensation at the $k=0$ mode. This alters the four-wave 
dynamics  and the system, after a strongly turbulent transient, becomes dominated by weak three-wave dynamics 
of acoustic  fluctuations  on background of  a strong coherent condensate. The long-time evolution in this
case is characterized 
by the appearance of a large quantum vortex ring in the numerical box. Kelvin waves propagate on this vortex ring,
 which causes the incompressible kinetic energy spectrum to follow the  $ -5/3 $ law. 
 We argue that the classical turbulence picture developed for the continuous vorticity fields is
 inapplicable; the fact that the Kelvin wave turbulence has the same spectrum as
the classical Kolmogorov spectrum is coincidental, as it arises from completely different physical
processes \cite{LN-spectrum,lvov-et-al-5/3}. The vortex ring in this regime coexists with the random
acoustic waves engaged in the three-wave interactions. The vortex shows up as a $v^{-3}$ tail on 
the velocity PDF while the random waves make up the Gaussian core of this PDF.

In {\bf RUN 2} and {\bf RUN 3} we have introduced a dissipation term at large scales
in order to stop the inverse cascade and to reach statistically steady states. 
The characteristics of these states depend strongly on the choice of the low-$k$ damping. 
If a friction is introduced ({\bf RUN 2}), the growth of the condensate is halted and the 
wave-action spectrum follows the four-wave WWT prediction. A large fuzzy vortex ring surrounded by 
  small ghost vortices appears in the final stage of the computation.
 If the dissipation is an hypo-viscosity ({\bf RUN 3}), the final steady spectra, 
 evaluated for a wide range of forcing coefficient, agree with the critical balance prediction. 
 In this regime the computational box appears to be filled by a vortex tangle - a chaotic set
 of strongly nonlinear
 vortex lines. The velocity PDF exhibits, both in the core and on the tails,
 a power-law behavior $v^{-3}$ characteristic to such vortex lines.
   





In Summary, our numerical results clearly show that  the turbulent state in the direct cascade range
 is strongly affected by the choice of damping in the inverse cascade range.
Most realistic configuration for the existing BEC experiments is configuration of {\bf RUN 1}. 
Indeed, while the hyper-viscosity can be physically understood as an evaporative cooling mechanism,
no large-scale damping mechanisms have ever been proposed. 
On the other hand, to study the nontrivial QT states predicted by the {\bf RUN 2} and the {\bf RUN 3} of the present paper,
it would be interesting to explore possibilities to damp the lowest-momentum modes in BEC experiments.

\section{Acknowledgments}
We thank Al Osborne and Victor L'vov for always stimulating discussions and for suggestions. We are also grateful to Guido Boffetta, Filippo De Lillo and Stefano Musacchio for precious advises on classical turbulence and numerics. We appreciate the work of FFTW developers in providing an excellent package to perform FFT algorithm.



\appendix

\section{Signs of the fluxes\label{ap:fluxes}}
The KZ solutions carry constant fluxes of the conserved quantities over the turbulent scales. 
The GPE model has two conserved quantities: the mass (particles) and the energy. 
Here, we will find the direction of the fluxes corresponding to the KZ solutions in a very simple way, 
 avoiding computing the flux based on a complicated relation resulting from the
  the kinetic equation, as it was done in \cite{kats1976} and \cite{dyachenko:1992hc}.

\begin{figure}
\includegraphics[scale=1.6]{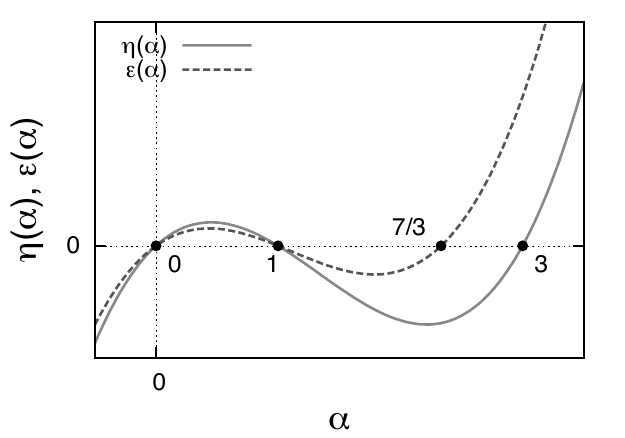}
\caption{Qualitative direction behaviors of energy and particles fluxes with respect to the power-law exponent $ \alpha $ of KZ solution $ n(k)=c k^{\alpha} $. \label{fig:sign_fluxes}}
\end{figure}

Consider Fig. \ref{fig:sign_fluxes}
where we qualitatively plot the flux on a generic power-law spectrum $ n(k)=c k^{-\alpha} $ (which is note necessarily a steady solution)
 as a function of the exponent $\alpha$. For very sharp spectra, $ \alpha \gg 1 $, both fluxes must
  be positive. To see that one can think of a narrow-band spectrum: it must spread which corresponds to the positive fluxes on the negative
   slope side. As the fluxes  are continuous functions of $\alpha$, we can determine their directions based on the zero-crossing 
   points, i.e. the RJ and the KZ exponents. 
    Of course, at the two thermodynamic RJ solutions,
 corresponding to $ \alpha=0 $ and $ \alpha=1 $, both fluxes must be zero.
 At the KZ solutions, only one of the fluxes turns into zero.
   Namely, on the energy cascade KZ, the mass flux is null and vice versa.
    Thus, the functions $ \eta(\alpha) $ and $ \epsilon(\alpha) $ qualitatively behave as 
    shown in Fig. \ref{fig:sign_fluxes}. It is then clear that the energy undergoes a direct cascade while the mass cascade inversely.

\section{Locality of interactions\label{ap:locality}}
Lets test the locality of interactions in the constant-flux states. This imply checkig that the collision integral in the kinetic equation (\ref{kineticequation}) converges on the KZ spectra $ n(k)=c k^{-\alpha} $. 
As these solutions are scale-invariant, the integral is easily written in the $ \omega $ space as
\begin{eqnarray}
I(\omega_1)= & \frac{c^3}{8 \omega_1^{\frac{\alpha}{2}}}\int (\omega_2 \omega_3 \omega_4)^{\frac{-\alpha+d-1}{2}} \left(\omega_1^{\frac{\alpha}{2}} + \omega_2^{\frac{\alpha}{2}} - \omega_3^{\frac{\alpha}{2}} -\omega_4^{\frac{\alpha}{2}}\right) \nonumber \\
& \times \delta(\omega_1+\omega_2-\omega_3-\omega_4) f(\omega_1, \omega_2, \omega_3, \omega_4) d\omega_{234},
\end{eqnarray}
where term
\begin{equation}
f(\omega_1, \omega_2, \omega_3, \omega_4)=\frac{\min\left[\sqrt{\omega_1}, \sqrt{\omega_2}, \sqrt{\omega_3},\sqrt{\omega_4}\right]}{2 (\omega_1 \omega_2 \omega_3 \omega_4)^{\frac{1}{2}}}
\end{equation}
takes into account the 3D average of $ \delta(\mathbf{k}_1+\mathbf{k}_2-\mathbf{k}_3-\mathbf{k}_4) $ over the
 solid angles $ \Omega_1, \Omega_2, \Omega_3$ and $\Omega_4 $ , see \cite{zakharov41kst, dyachenko:1992hc, Connaughton200491, pon:boltzm-dam} for details. In this coordinate system the frequency $ \delta $-function can be 
 easily used, for example as $ \omega_2 =\omega_3+\omega_4-\omega_1 $. The integral presents singularities for integration over $ \omega_3 $ at zero and infinity.
Note that is the same is true  for $ \omega_4 $ integration as the integral is symmetric with respect to
 $ (3, 4)\rightarrow(4, 3) $. By Taylor expanding the integrand  up to the leading  order, one gets
\begin{equation}
\begin{split}
& \lim_{\omega_3\rightarrow\infty} I(\omega_1) \sim \int \omega_3^{-\frac{\alpha+2}{2}} d\omega_3 \;\;\; \mbox{and} \\
& \lim_{\omega_3\rightarrow\ 0} I(\omega_1) \sim \int \omega_3^{-\frac{\alpha-1}{2}} d\omega_3.
\end{split}
\end{equation}
The locality holds for $ 0 < \alpha < 3 $ which is true for the inverse particles cascade ($ \alpha=7/3 $) but not for the direct energy cascade ($ \alpha=3 $). Nevertheless the divergence in the latter case is marginal
 and the solution can be corrected by a logarithmic factor \cite{dyachenko:1992hc}.

\section{The Fj{\o}rtoft argument}\label{fjortoft}
We will present here a new version of the  argument Fj{\o}rtoft argument which 
is formulated for the conservative systems (no forcing or dissipation).
Let us introduce the mass and the energy
{\it centroids}  as
\begin{subequations}
\begin{equation}
K_M=\frac{\int k n(k) dk}{M}
\end{equation}
\begin{equation}
K_E=\frac{\int k \mathcal{E}(k) dk}{E}=\frac{\int k^3 n(k) dk}{E}.
\end{equation}
\end{subequations}
In the latter expression we have assumed (this is essential for the  Fj{\o}rtoft argument) that 
the nonlinear energy is negligible with respect to linear one.
In the following will use the Cauchy-Schwartz inequality
\begin{equation}
\int f(k) g(k) dk \le \left( \int f^2(k) dk \right)^{1/2} \left( \int g^2(k) dk \right)^{1/2}
\end{equation}
By splitting the  energy integrand in two parts we get
\begin{eqnarray}
E & = & \int k^2 n(k) dk = \int (k^{1/2} n^{1/2}) \times (k^{3/2} n^{1/2}) dk \nonumber \\
& \le & \left( \int k n(k) dk \right)^{1/2} \left( \int k^3 n(k) dk \right)^{1/2} = \sqrt{K_M M K_E E}
\end{eqnarray}
and so we have 
\begin{equation}
\label{eq;cond}
 K_M K_E \ge \frac{E}{M}.
\end{equation}
This inequality means that if the particle centroid moves to low wave numbers (inverse cascade)
the energy centroid must move to high wave numbers (direct cascade).
 
 We now evaluate
\begin{eqnarray}
K_M M & = & \int k n(k) dk = \int (k n^{1/2}) \times (n^{1/2}) dk \nonumber \\
& \le & \left( \int k^2 n(k) dk \right)^{1/2} \left( \int n(k) dk \right)^{1/2} = \sqrt{M E}
\end{eqnarray}
which gives 
\begin{equation}
\label{eq;Mcasc}
K_M \le \sqrt{E/M}.
\end{equation}
This inequality means that  the mass centroid can either stay where it is initially, or move
to the large scales (inverse cascade), but it cannot cascade to the small scales.
 
 Combining the inequalities \eqref{eq;cond} and \eqref{eq;Mcasc}, we get
 \begin{equation}
 \label{eq;cond1}
 K_E \ge \sqrt{E/M}.
\end{equation}
This inequality means that if the energy centroid can either stay where it was initially, or move
to the smaller scales (direct cascade), but it cannot cascade to the larger scales.



\bibliographystyle{elsarticle-num}
\bibliography{references}

\begin{thebibliography}{10}
\expandafter\ifx\csname url\endcsname\relax
  \def\url#1{\texttt{#1}}\fi
\expandafter\ifx\csname urlprefix\endcsname\relax\def\urlprefix{URL }\fi
\expandafter\ifx\csname href\endcsname\relax
  \def\href#1#2{#2} \def\path#1{#1}\fi

\bibitem{frisch1995t}
U.~Frisch, {Turbulence: the legacy of AN Kolmogorov}, Cambridge University
  Press, 1995.

\bibitem{zakh65}
V.~Zakharov, {Weak turbulence in media with decay spectrum,}, J. Appl. Mech.
  Tech. Phys.~(4) (1965) 22--24.

\bibitem{zakharov41kst}
V.~Zakharov, V.~L'vov, G.~Falkovich, {Kolmogorov Spectra of Turbulence 1: Wave
  Turbulence}, Springer-Verlag, 1992.

\bibitem{zakharov1967ess}
V.~Zakharov, N.~Filonenko, {Energy Spectrum for Stochastic Oscillations of the
  Surface of a Liquid}, in: Soviet Physics Doklady, Vol.~11, 1967, p. 881.

\bibitem{nazarenko-book}
S.~Nazarenko, {Wave Turbulence}, Springer-Verlag, 2010.

\bibitem{sulem1899nse}
C.~Sulem, P.~Sulem, {The Nonlinear Schr{\"o}dinger Equation: Self-Focusing and
  Wave Collapse}, Springer, 1899.

\bibitem{dalfovo1999theory}
F.~Dalfovo, S.~Giorgini, L.~Pitaevskii, S.~Stringari, {Theory of Bose-Einstein
  condensation in trapped gases}, Reviews of Modern Physics 71~(3) (1999)
  463--512.

\bibitem{koplik1993vrs}
J.~Koplik, H.~Levine, {Vortex reconnection in superfluid helium}, Physical
  Review Letters 71~(9) (1993) 1375--1378.

\bibitem{berloff2002ssn}
N.~Berloff, B.~Svistunov, {Scenario of strongly nonequilibrated Bose-Einstein
  condensation}, Physical Review A 66~(1) (2002) 13603.

\bibitem{nore1997dkt}
C.~Nore, M.~Abid, M.~Brachet, {Decaying Kolmogorov turbulence in a model of
  superflow}, Physics of Fluids 9 (1997) 2644.

\bibitem{kobayashi2005kss}
M.~Kobayashi, M.~Tsubota, {Kolmogorov Spectrum of Superfluid Turbulence:
  Numerical Analysis of the Gross-Pitaevskii Equation with a Small-Scale
  Dissipation}, Physical Review Letters 94~(6) (2005) 65302.

\bibitem{kobayashi2005kol}
M.~Kobayashi, M.~Tsubota, Kolmogorov spectrum of quantum turbulence,
  JOURNAL-PHYSICAL SOCIETY OF JAPAN 74~(12) (2005) 3248.

\bibitem{proment:051603}
D.~Proment, S.~Nazarenko, M.~Onorato,
  \href{http://link.aps.org/abstract/PRA/v80/e051603}{Quantum turbulence
  cascades in the gross-pitaevskii model}, Physical Review A (Atomic,
  Molecular, and Optical Physics) 80~(5) (2009) 051603.
\newblock \href {http://dx.doi.org/10.1103/PhysRevA.80.051603}
  {\path{doi:10.1103/PhysRevA.80.051603}}.
\newline\urlprefix\url{http://link.aps.org/abstract/PRA/v80/e051603}

\bibitem{kozik:060502}
E.~Kozik, B.~Svistunov,
  \href{http://link.aps.org/abstract/PRB/v77/e060502}{Kolmogorov and
  kelvin-wave cascades of superfluid turbulence at t = 0: What lies between},
  Physical Review B (Condensed Matter and Materials Physics) 77~(6) (2008)
  060502.
\newblock \href {http://dx.doi.org/10.1103/PhysRevB.77.060502}
  {\path{doi:10.1103/PhysRevB.77.060502}}.
\newline\urlprefix\url{http://link.aps.org/abstract/PRB/v77/e060502}

\bibitem{alamri:215302}
S.~Z. Alamri, A.~J. Youd, C.~F. Barenghi,
  \href{http://link.aps.org/abstract/PRL/v101/e215302}{Reconnection of
  superfluid vortex bundles}, Physical Review Letters 101~(21) (2008) 215302.
\newblock \href {http://dx.doi.org/10.1103/PhysRevLett.101.215302}
  {\path{doi:10.1103/PhysRevLett.101.215302}}.
\newline\urlprefix\url{http://link.aps.org/abstract/PRL/v101/e215302}

\bibitem{leadbeater2001sou}
M.~Leadbeater, T.~Winiecki, D.~Samuels, C.~Barenghi, C.~Adams, Sound emission
  due to superfluid vortex reconnections, Physical Review Letters 86~(8) (2001)
  1410--1413.

\bibitem{url:proment-qt}
D.~Proment, S.~Nazarenko, M.~Onorato,
  \href{http://www.youtube.com/watch?v=uk5DpF4vnFs}{Quantum vortex decay
  (movie)}.
\newline\urlprefix\url{http://www.youtube.com/watch?v=uk5DpF4vnFs}

\bibitem{vinen-nimela}
W.~F. Vinen, J.~Niemela, Journal of Low Temperature Physics 128 (2002) 167.

\bibitem{PhysRevB.61.1410}
W.~F. Vinen, Classical character of turbulence in a quantum liquid, Phys. Rev.
  B 61~(2) (2000) 1410--1420.
\newblock \href {http://dx.doi.org/10.1103/PhysRevB.61.1410}
  {\path{doi:10.1103/PhysRevB.61.1410}}.

\bibitem{PhysRevB.64.134520}
W.~F. Vinen, Decay of superfluid turbulence at a very low temperature: The
  radiation of sound from a kelvin wave on a quantized vortex, Phys. Rev. B
  64~(13) (2001) 134520.
\newblock \href {http://dx.doi.org/10.1103/PhysRevB.64.134520}
  {\path{doi:10.1103/PhysRevB.64.134520}}.

\bibitem{lvov_naz_rud1}
V.~L'vov, S.~Nazarenko, O.~Rudenko, { Bottleneck crossover between classical
  and quantum superfluid turbulence}, Phys Rev B 76~(2) (2007) 024520.

\bibitem{lvov_naz_rud2}
V.~L'vov, S.~Nazarenko, O.~Rudenko, { Gradual eddy-wave crossover in superfluid
  turbulence}, Journal of Low Temperature Physics 153~(5-6) (2008) 140--161.

\bibitem{PhysRevLett.92.035301}
E.~Kozik, B.~Svistunov, Kelvin-wave cascade and decay of superfluid turbulence,
  Phys. Rev. Lett. 92~(3) (2004) 035301.
\newblock \href {http://dx.doi.org/10.1103/PhysRevLett.92.035301}
  {\path{doi:10.1103/PhysRevLett.92.035301}}.

\bibitem{boffetta}
G.~Boffetta, A.~Celani, D.~Dezzani, J.~Laurie, S.~Nazarenko, {Modeling Kelvin
  Wave Cascades in Superfluid Helium}, Journal of Low Temperature Physics
  156~(3-6) (2009) 193--214.

\bibitem{KW_rec}
S.~Nazarenko, {Kelvin wave turbulence generated by vortex reconnections}, JETP
  Letters 84~(11) (2006) 585--587.

\bibitem{KW-dam}
S.~Nazarenko, {Differential approximation for Kelvin-wave turbulence}, JETP
  Letters 83~(5) (2005) 198--200.

\bibitem{LN-spectrum}
V.~L'vov, S.~Nazarenko, {Spectrum of Kelvin-wave turbulence in superfluids},
  JETP Letters 91~((8)) (2010) 464--470.

\bibitem{yepez:084501}
J.~Yepez, G.~Vahala, L.~Vahala, M.~Soe,
  \href{http://link.aps.org/abstract/PRL/v103/e084501}{Superfluid turbulence
  from quantum kelvin wave to classical kolmogorov cascades}, Physical Review
  Letters 103~(8) (2009) 084501.
\newblock \href {http://dx.doi.org/10.1103/PhysRevLett.103.084501}
  {\path{doi:10.1103/PhysRevLett.103.084501}}.
\newline\urlprefix\url{http://link.aps.org/abstract/PRL/v103/e084501}

\bibitem{PhysRevLett.104.075301}
A.~C. White, C.~F. Barenghi, N.~P. Proukakis, A.~J. Youd, D.~H. Wacks,
  Nonclassical velocity statistics in a turbulent atomic bose-einstein
  condensate, Phys. Rev. Lett. 104~(7) (2010) 075301.
\newblock \href {http://dx.doi.org/10.1103/PhysRevLett.104.075301}
  {\path{doi:10.1103/PhysRevLett.104.075301}}.

\bibitem{cln0}
Y.~Choi, Y.~Lvov, S.~Nazarenko, P.~B., {Anomalous probability of large
  amplitudes in wave turbulence}, Phys. Lett. A 339 (2005) 361--369.

\bibitem{cln1}
Y.~Choi, Y.~Lvov, S.~Nazarenko, {Probability densities and preservation of
  randomness in wave turbulence}, Phys. Lett. A 332 (2004) 230--238.

\bibitem{cln2}
Y.~Choi, Y.~Lvov, S.~Nazarenko, {Joint statistics of amplitudes and phases in
  wave turbulence}, Physica D 201 (2005) 121--149.

\bibitem{dyachenko:1992hc}
S.~Dyachenko, A.~C. Newell, A.~Pushkarev, V.~E. Zakharov,
  \href{http://www.sciencedirect.com/science/article/B6TVK-46JH21H-4G/2/b9bf3a%
47086f6f154a8c0478ca64c07b}{Optical turbulence: weak turbulence, condensates
  and collapsing filaments in the nonlinear schr{\"o}dinger equation}, Physica
  D: Nonlinear Phenomena 57~(1-2) (1992) 96--160.
\newline\urlprefix\url{http://www.sciencedirect.com/science/article/B6TVK-46JH%
21H-4G/2/b9bf3a47086f6f154a8c0478ca64c07b}

\bibitem{PhysRevLett.95.263901}
C.~Connaughton, C.~Josserand, A.~Picozzi, Y.~Pomeau, S.~Rica, Condensation of
  classical nonlinear waves, Phys. Rev. Lett. 95~(26) (2005) 263901.
\newblock \href {http://dx.doi.org/10.1103/PhysRevLett.95.263901}
  {\path{doi:10.1103/PhysRevLett.95.263901}}.

\bibitem{connaughton2003dimensional}
C.~Connaughton, S.~Nazarenko, A.~Newell, {Dimensional analysis and weak
  turbulence}, Physica D: Nonlinear Phenomena 184~(1-4) (2003) 86--97.

\bibitem{zakharov2005dbe}
V.~Zakharov, S.~Nazarenko, {Dynamics of the Bose--Einstein condensation},
  Physica D: Nonlinear Phenomena 201~(3-4) (2005) 203--211.

\bibitem{zakharov1970sat}
V.~Zakharov, R.~Sagdeev, {Spectrum of acoustic turbulence}, Soviet Physics -
  Doklady 15~(4).

\bibitem{goldreich1995tti}
P.~Goldreich, S.~Sridhar, {Toward a theory of interstellar turbulence. 2:
  Strong alfvenic turbulence}, The Astrophysical Journal 438 (1995) 763--775.

\bibitem{nazarenko-cb-paper}
S.~Nazarenko, A.~Schekochihin, {Critical balance in magnetohydrodynamic,
  rotating and stratified turbulence: towards a universal scaling conjecture},
  J. Fluid Mech. (submitted).

\bibitem{phillips:367511}
O.~M. Phillips, The equilibrium range in the spectrum of wind-generated waves,
  Journal of Fluid Mechanics Digital Archive 4~(04) (1958) 426--434.
\newblock \href {http://dx.doi.org/10.1017/S0022112058000550}
  {\path{doi:10.1017/S0022112058000550}}.

\bibitem{korotkevich:074504}
A.~O. Korotkevich,
  \href{http://link.aps.org/abstract/PRL/v101/e074504}{Simultaneous numerical
  simulation of direct and inverse cascades in wave turbulence}, Physical
  Review Letters 101~(7) (2008) 074504.
\newblock \href {http://dx.doi.org/10.1103/PhysRevLett.101.074504}
  {\path{doi:10.1103/PhysRevLett.101.074504}}.
\newline\urlprefix\url{http://link.aps.org/abstract/PRL/v101/e074504}

\bibitem{press:1992rm}
W.~Press, B.~Flannery, S.~Teukolsky, W.~Vetterling,
  \href{http://www.amazon.ca/exec/obidos/redirect?tag=citeulike09-20&amp;path=%
ASIN/0521431085}{Numerical Recipes in C : The Art of Scientific Computing},
  {\{}Cambridge University Press{\}}, 1992.
\newline\urlprefix\url{http://www.amazon.ca/exec/obidos/redirect?tag=citeulike%
09-20&amp;path=ASIN/0521431085}

\bibitem{url:fftw}
\href{http://www.fftw.org/}{Fast fourier transform of the west}.
\newline\urlprefix\url{http://www.fftw.org/}

\bibitem{bao:2006zk}
W.~Bao, H.~Wang,
  \href{http://www.sciencedirect.com/science/article/B6WHY-4J9X1W0-2/2/141446f%
cf67a4f984b0eddb82e4d0a52}{An efficient and spectrally accurate numerical
  method for computing dynamics of rotating bose-einstein condensates}, Journal
  of Computational Physics 217~(2) (2006) 612--626.
\newline\urlprefix\url{http://www.sciencedirect.com/science/article/B6WHY-4J9X%
1W0-2/2/141446fcf67a4f984b0eddb82e4d0a52}

\bibitem{nazarenko:2006ta}
S.~Nazarenko, Sandpile behaviour in discrete water-wave turbulence, Journal of
  Statistical Mechanics: Theory and Experiment 2006~(02) (2006) L02002--L02002.

\bibitem{zakharov:2005jt}
V.~Zakharov, A.~Korotkevich, A.~Pushkarev, A.~Dyachenko,
  \href{http://dx.doi.org/10.1134/1.2150867}{Mesoscopic wave turbulence}, JETP
  Letters 82~(8) (2005) 487--491.
\newline\urlprefix\url{http://dx.doi.org/10.1134/1.2150867}

\bibitem{falkovich1988efc}
G.~Falkovich, A.~Shafarenko, {What energy flux is carried away by the
  Kolmogorov weak turbulence spectrum?}, Soviet Physics - JETP 68~(1) (1988)
  1393--1397.

\bibitem{paoletti:154501}
M.~S. Paoletti, M.~E. Fisher, K.~R. Sreenivasan, D.~P. Lathrop,
  \href{http://link.aps.org/abstract/PRL/v101/e154501}{Velocity statistics
  distinguish quantum turbulence from classical turbulence}, Physical Review
  Letters 101~(15) (2008) 154501.
\newblock \href {http://dx.doi.org/10.1103/PhysRevLett.101.154501}
  {\path{doi:10.1103/PhysRevLett.101.154501}}.
\newline\urlprefix\url{http://link.aps.org/abstract/PRL/v101/e154501}

\bibitem{lvov-et-al-5/3}
V.~L'vov, S.~Nazarenko, D.~Proment, Kelvin-wave interpretation of the $ -5/3 $
  spectrum in superfluid turbulence, in preparation (2010).

\bibitem{kats1976}
A.~Kats, {Direction of transfer of energy and quasi-particle number along the
  spectrum in stationary power-law solutions of the kinetic equations for waves
  and particles}, Soviet Journal of Experimental and Theoretical Physics 44
  (1976) 1106.

\bibitem{Connaughton200491}
C.~Connaughton, Y.~Pomeau,
  \href{http://www.sciencedirect.com/science/article/B6X19-4BRTCTX-2/2/40a1ed5%
455ce5eb762bcb918148dd8a4}{Kinetic theory and bose-einstein condensation},
  Comptes Rendus Physique 5~(1) (2004) 91 -- 106, bose-Einstein condensates:
  recent advances in collective effects.
\newblock \href {http://dx.doi.org/DOI: 10.1016/j.crhy.2004.01.006}
  {\path{doi:DOI: 10.1016/j.crhy.2004.01.006}}.
\newline\urlprefix\url{http://www.sciencedirect.com/science/article/B6X19-4BRT%
CTX-2/2/40a1ed5455ce5eb762bcb918148dd8a4}

\bibitem{pon:boltzm-dam}
D.~Proment, M.~Onorato, P.~Asinari, S.~Nazarenko, Turbulent cascades in the
  boltzmann kinetic equation, in preparation.

\end{thebibliography}






\end{document}